\documentclass[10pt,pre,twocolumn,aps,showpacs,superscriptaddress]{revtex4-1}
\usepackage{epsfig,amsmath,amssymb,graphicx,color,calc,epstopdf}

\let\a=\alpha  \let\g=\gamma \let\d=\delta
\let\e=\varepsilon  \let\h=\eta \let\k=\kappa
\let\l=\lambda \let\m=\mu   
\let\s=\sigma

\let\ee=\epsilon \let\r=\rho  \let\io=\infty

 \def\HH{{\cal H}}
\def\NN{{\cal N}} 
  \def\OO{{\cal O}}
 \def\SS{{\cal S}}

\def\to{\rightarrow} \def\la{\left\langle} \def\ra{\right\rangle}

\newcommand{\Tr}{\text{Tr}}

\newcommand{\beq}{\begin{equation}}
\newcommand{\eeq}{\end{equation}}
\newcommand{\ba}{\begin{align}}
\newcommand{\ea}{\end{align}}

\begin{document}

\title{Jamming Criticality Revealed by Removing Localized Buckling Excitations}

\author{Patrick Charbonneau}
\affiliation{Department of Chemistry, Duke University, Durham,
North Carolina 27708, USA}
\affiliation{Department of Physics, Duke University, Durham,
North Carolina 27708, USA}

\author{Eric I. Corwin}
\affiliation{Department of Physics, University of Oregon, Eugene, Oregon 97403, USA}

\author{Giorgio Parisi}
\affiliation{Dipartimento di Fisica,
Sapienza Universit\`a di Roma,
INFN, Sezione di Roma I, IPFC -- CNR,
Piazzale Aldo Moro 2, I-00185 Roma, Italy
}

\author{Francesco Zamponi}
\affiliation{LPT,
\'Ecole Normale Sup\'erieure, UMR 8549 CNRS, 24 Rue Lhomond, 75005 France}

\begin{abstract}
Recent theoretical advances offer an exact, first-principle theory of jamming criticality in infinite dimension as well as universal scaling relations between critical exponents in all dimensions. 
For packings of frictionless spheres near the jamming transition, these advances predict that
nontrivial power-law exponents characterize the critical distribution of {\it (i)} small inter-particle gaps and {\it (ii)} weak contact forces, both of which are crucial for mechanical stability.
The scaling of the inter-particle gaps is known to be constant in all spatial dimensions $d$ -- including the physically relevant $d=2$ and 3, but
the value of the weak force exponent remains the object of debate and confusion. 
Here, we resolve this ambiguity by numerical simulations. We construct isostatic jammed packings with extremely high accuracy,
and introduce a simple criterion to separate 
the contribution of particles that give rise to localized buckling excitations, i.e., \emph{bucklers}, from the others. 
This analysis reveals the remarkable dimensional robustness of mean-field marginality and its associated criticality.
\end{abstract}


\maketitle
\begin{figure*}
\center{\includegraphics[width=0.7\columnwidth]{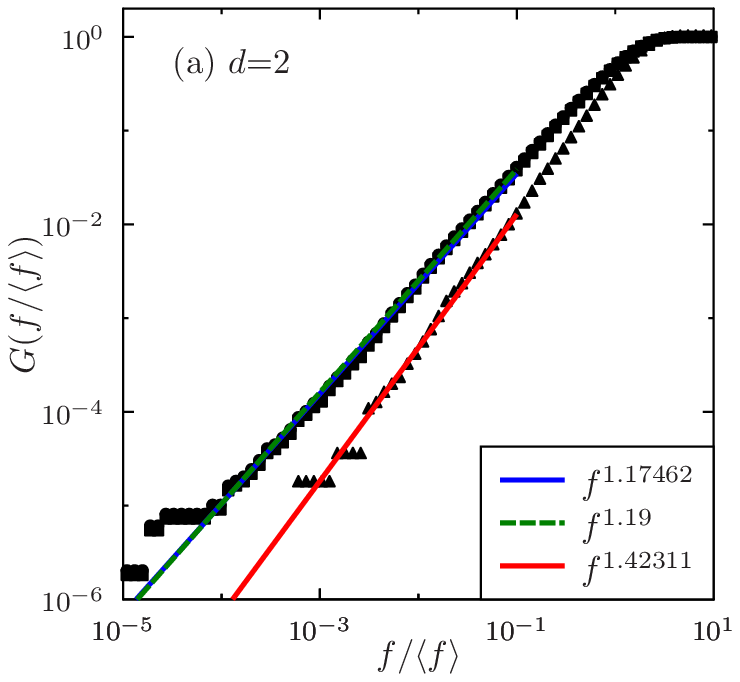}\includegraphics[width=0.7\columnwidth]{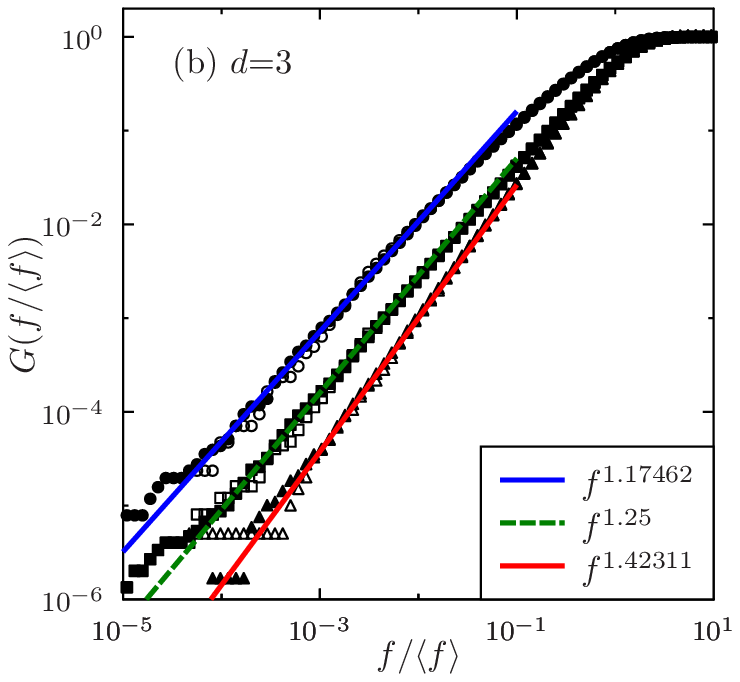}\includegraphics[width=0.7\columnwidth]{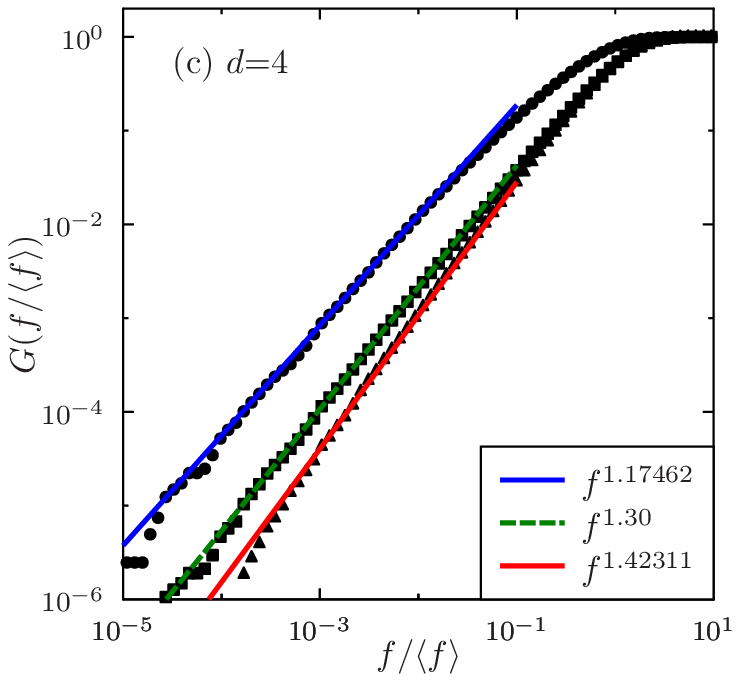}}
\caption{Cumulative force distribution $G(f)$ for $d$= 2, 3, and 4, in (a), (b), and (c), respectively. The distributions $P(f)$ (square), $P_\ell(f)$ (circle), and $P_{\mathrm{e}}(f)$ (triangle) all show power-law scalings at small forces (in $d=2$, squares and circles are nearly superimposed because of the large proportion of bucklers), but with different exponents. For the combined distribution, $\theta_{\mathrm f}$ grows steadily with $d$, but $P_\ell(f)$ and $P_{\mathrm{e}}(f)$ are consistent with the exponents obtained by combining the $d=\infty$ solution with 
marginal stability arguments, i.e., $\theta_{\ell}=0.17462$ and $\theta_{\mathrm{e}}=0.42311$, respectively. In $d=2$, the results for $\theta_{\mathrm{e}}$ are consistent with those obtained by ranking contacts based on the coupling of their related soft modes to external forces~\cite{LDW13}. In $d=3$, comparing two system sizes, $N_{\mathrm{i}}=16384$ (full symbols) and $N_{\mathrm{i}}=1024$ (empty symbols), confirms the absence of significant finite-size effects. }
\label{fig:forces}
\end{figure*}

\section{Introduction} 
The analogy between glasses and sand piles, which are both rigid and disordered,
was first proposed by Bernal~\cite{Be59},
who pioneered comparing 
their structure. The analogy received 
renewed attention following Liu and Nagel's suggestion
that different disordered solids could be described by the same 
jamming phase diagram~\cite{LN98}. 
Motivated by the ubiquity of jamming in materials physics, an intense research effort from the 
soft and granular matter community, on the one hand, and from 
the statistical mechanics of disordered systems community, 
on the other, has since ensued~\cite{He10,TS10,PZ10}.

Recent theoretical breakthroughs have succeeded in 
transforming this analogy into a solid predictive 
framework. Quite remarkably, results from what appeared, at first, to be two independent 
lines of work now point towards a unifying view of the 
glass problem, understood as encompassing 
a broad range of amorphous materials. 
First, the \emph{exact} infinite-dimensional (mean-field) solution of the celebrated hard-sphere model 
precisely unifies glass formation and jamming~\cite{PZ10,KPZ12,KPUZ13,CKPUZ14,CKPUZ14b}. 
This $d=\infty$ solution predicts, from first principles, that jammed packings are mechanically stable but only marginally so~\cite{CKPUZ14,CKPUZ14b}. The packings are therefore isostatic~\cite[Sec.~A]{foot:2}, i.e., the number of interparticle contacts  matches Maxwell's criterion for mechanical stability~\cite{He10,TS10}. 
The solution further predicts a (non-trivial) criticality near jamming~\cite{CKPUZ14,CKPUZ14b}. 
Second, a real-space description of elementary excitations near jamming finds that soft modes pervade in that regime~\cite{WSNW05,Brito09}. This approach further provides scaling relations between the jamming critical exponents, based on marginal stability~\cite{Wy12,LDW13,DeGiuli14,DLBW14,MW14}.

Both approaches agree that two power-law exponents characterize the structure of disordered jammed packings. The distribution 
{\it (i)} of spatial gaps between particles that are nearly (but not quite) in contact, and hence of the average number of 
neighbors away from a sphere surface, scales as $Z(h)-Z(0) \sim h^{1-\gamma}$ for small $h$, where $h = (r - \sigma)/\s$  is the
gap size for spheres of diameter $\sigma$~\cite{foot:1}; and {\it (ii)} of weak forces $f$ between spheres scales as $P(f) \sim f^{\theta_\mathrm{f}}$ for small $f$.
The $d=\infty$ solution further predicts $\gamma=0.41269\ldots$ and
$\theta_{\mathrm{f}}=0.42311\ldots$~\cite{CKPUZ14,CKPUZ14b}.

For isostatic jammed packings, it is found that opening a force contact between particles destabilizes the system by creating a soft mode~\cite[Sec.~A7]{foot:2},
which is a collective particle excitation that preserves the remaining contacts~\cite{Wy12,LDW13,LDDW14,DLBW14}. 
Phenomenologically, it was noted that some of the resulting excitations are extended while others are localized~\cite{LDW13}. 
This distinction suggests the existence of two different force exponents: $P(f) \sim f^{\theta_\mathrm{e}}$ for contacts associated with extended modes, 
and $P(f) \sim f^{\theta_\ell}$ for contacts associated with localized modes. The
observed total force distribution, which is a weighted average of the two, 
should therefore have the asymptotic form $P(f) \sim f^{\theta_\mathrm{f}}$, with $\theta_\mathrm{f} =\mathrm{min}(\theta_\mathrm{e}, \theta_\ell)$. Marginal mechanical stability analysis provides universal scaling relations for the exponents: $\theta_\ell=1-2\gamma$ and $\theta_\mathrm{e}=\theta_\ell/\gamma$~\cite{Wy12,LDW13,DeGiuli14,DLBW14}. 

We are now, however, left with a conundrum. Using $\gamma=0.41269$ from the $d=\io$ solution with the scaling relations derived from marginal stability gives
$\theta_{\mathrm{e}}=0.42311$ and $\theta_\ell = 0.17462$.
The $d=\infty$ solution thus \emph{exactly} obeys the scaling relations, 
but only if one assumes $\theta_\mathrm{f}=\theta_{\mathrm{e}}$. 
Yet this assumption is inconsistent with the relation $\theta_\mathrm{f} =\mathrm{min}(\theta_\mathrm{e}, \theta_\ell) = \theta_\ell$, 
which must be true if localized modes exist with finite probability.

The situation is further muddled by the currently available numerical estimates for $\gamma$ and $\theta_\mathrm{f}$ in finite $d$.
The value $\gamma=0.40(2)$ is found to remain unchanged for all $d\geq 2$~\cite{SDST06,GLN12,CCPZ12,LDW13,DLBW14}, 
and is consistent with the $d=\infty$ solution. The upper critical dimension for the jamming transition 
may thus be as low as $d_\mathrm{u}=2$~\cite{GLN12,CCPZ12}, with all jamming critical exponents being constant for $d\geq2$. 
Reported values for $\theta_\mathrm{f}$, however, 
range from 0~\cite{OLLN02,DTS05} to 0.42~\cite{CCPZ12}. Encouragingly, for $d=2$ the most reliable determinations found $\theta_\mathrm{f}=0.18(2)$~\cite{LDW13,DLBW14}, independently of the interaction potential~\cite[Sec.~A8]{foot:2}. 
For $d$=3, however, no such agreement is observed and the results even depend on microscopic details of the system~\cite{DLBW14}. In this letter, we resolve this perplexing situation by identifying a simple geometrical criterion associated with localized modes,
which allows us to accurately study the weak tail of the force distribution.

\section{Isostaticity and the force network}
Consider a packing with $i = 1\cdots N$ particles located in positions $\mathbf{r}_i = \{ r_{i\a} \}$
in $\alpha =1 \cdots d$ dimensions with $\la ij \ra = 1\cdots N_{\mathrm{c}}$ contacts, where $i<j$. 
At the jamming transition, particles do not overlap, hence $|\mathbf{r}_j - \mathbf{r}_i| \geq \sigma_{ij}$, where $\sigma_{ij}=(\sigma_i+\sigma_j)/2$ is the sum of particle radii.
Two particles are in contact if $|\mathbf{r}_j - \mathbf{r}_i| = \sigma_{ij}$, and in this case they exchange a radial force
along the contact vector $\mathbf{n}_{ij} = (\mathbf{r}_j - \mathbf{r}_i)/|\mathbf{r}_j - \mathbf{r}_i|$. 
The scalar contact force $f_{ij} = f_{ji}$ on each contact defines $\vec f = \{ f_{ij} \}$, an $N_{\mathrm{c}}$-dimensional vector, and
the external forces $F_{i\alpha}$ define
an $Nd$-dimensional vector $\vec F = \{ F_{i\alpha} \}$.
The force balance equations for particle $i$, given the set $\partial i$ of particles in contact with it, then reads
\beq\label{forcebalance}
F_{i\alpha}  = \sum_{j \in \partial i} n_{ji}^\alpha f_{ji} \ \ \ \ \ \
\Rightarrow
\ \ \ \ \ \ 
\vec F = \SS^T \vec f \ ,
\eeq
where $\SS$ is a $N_{\mathrm{c}} \times Nd$ matrix with elements $\SS_{\la ij \ra}^{k \alpha} = ( \d_{jk} - \d_{ik}) n_{ij}^\alpha$~\cite[Sec.~A1]{foot:2}.
For a system under cubic periodic boundary conditions and in mechanical equilibrium under no external force, i.e., $\vec F=\vec 0$,
Eq.~\eqref{forcebalance} 
gives $Nd$ \emph{homogeneous} linear equations for the $N_{\mathrm{c}}$ contact forces, i.e., $\SS^T \vec f =0$.
The contact vector $\vec f$ is therefore a zero mode of $\SS^T$.
For convenience, we define the $N_{\mathrm{c}} \times N_{\mathrm{c}}$ symmetric matrix $\NN = \SS \, \SS^T$, 
which has all the zero modes of $\SS^T$, but may also have additional ones~\cite[Sec.~A3]{foot:2}. 

After taking into account global translational invariance, Equation~\eqref{forcebalance} results in a system of $(N-1)d$ homogeneous linearly independent equations over $N_{\mathrm{c}}$ variables~\cite[Sec.~A1]{foot:2}, and
therefore admits $\max\{N_{\mathrm{c}} - (N-1)d, 0\}$ non-zero linearly independent solutions. 
It has further been argued that jamming takes place in the isostatic limit~\cite{Wy12,GLN12,HST13,LDW13}, 
which corresponds to the existence of a single solution to $\SS^T\vec{f}=0$ (Eq.~\eqref{forcebalance}), and hence $N_{\mathrm{c}}=(N-1)d+1$ under periodic boundary 
conditions~\cite[Secs.~A6 and B1]{foot:2}. Note that in the thermodynamic limit the average particle connectivity, $Z(0) = 2N_{\mathrm{c}}/N$, 
converges to the usual Maxwell criterion for mechanical stability, $\lim_{N\rightarrow\infty}Z(0)=2d$, consistently with the $d=\io$ solution~\cite[Sec.~B1]{foot:2}.
For an isostatic system, $\NN$ has a unique zero mode~\cite[Sec.~A6]{foot:2}, and
because $\NN = \SS \SS^T$ we also have $\NN\vec{f}=0$. The solution vector $\vec{f}$ must therefore be the unique zero mode of $\NN$, 
with an overall scale factor corresponding to the global pressure. 
In summary, given the orientation vectors for each pair of contacts in an isostatic packing, the distribution of contact forces can be uniquely determined by finding the eigenvector corresponding to the zero-eigenvalue of $\NN$. 

\section{Numerical construction of jammed packings and calculation of the forces} 
Several protocols have been proposed to construct jammed packings of frictionless spheres, see, e.g.~\cite{DTS05,TJ10,OLLN02,GLN12,CCPZ12,HST13,LDW13}. 
Some of them, however, do not systematically result in packings that are precisely isostatic~\cite[Sec.~B1]{foot:2}. Because the scaling laws between the jamming exponents follow from isostaticity~\cite{Wy12,DeGiuli14,DLBW14}, this requirement is here strictly enforced~\cite[Sec.~B2]{foot:2}. Isostatic packings under periodic boundary conditions are numerically obtained by minimizing the energy of athermal soft spheres with a quadratic contact potential on general purpose graphical processing units using quad-precision calculations~\cite{CCPZ12,morse2014geometric,NVIDIA}.  
Our protocol begins with $N_{\mathrm{i}}$ randomly-distributed particles at a packing fraction $\varphi$ that is roughly twice the final jamming density $\varphi_{\mathrm{J}}$.  An isostatic point is approached by successively minimizing the system potential energy $U$ using a FIRE algorithm \cite{bitzek2006structural}, and then shrinking the particle radii.  The isostatic configuration can be efficiently approached by exploiting the scaling $U \propto \left( \varphi-\varphi_{\mathrm{J}} \right)^2$~\cite{OLLN02,JBZ11,CCPZ12} to iteratively estimate the value of $\varphi_{\mathrm{J}}$, and then target a new density at an energy that is a fixed fraction of the previous one~\cite[Sec.~B2]{foot:2}.  
For $d=3$ and 4, we thereby obtained approximately 100 single-component systems with $N_{\mathrm{i}}=16384$, and 100 equimolar binary mixtures with $N_{\mathrm{i}}=4096$ in $d=2$ with a diameter ratio of 1:1.4. A more limited set of configurations was also obtained for $N_{\mathrm{i}}=4096$ in $d=5-8$. In all cases, the choice of system and preparation protocols are known to fully suppress crystallization~\cite{perera:1998,SDST06,VCFC09,CCPZ12}. Note that applying this same preparation protocol to any other contact potential form (e.g., Hertzian) would also result in configurations that are valid hard sphere packings~\cite[Sec.~A]{foot:2}. Although the specific packing sensitively depends on the choice of contact potential, algorithm details and initial configuration, similar structural scaling relations are known to be robustly independent of this choice~\cite{CCPZ12,LDW13,DLBW14}.

We analyze the contact network at $\varphi_{\mathrm{J}}$ by first eliminating particles with $Z<d+1$ contacts, i.e., the {\it rattlers} (Fig.~\ref{fig:sketch})~\cite{DTS05,OLLN02}.
After this step, most configurations have $N_{\mathrm{c}} = (N-1)d+1$, where $N$ is the number of remaining particles. We discard the configurations for which this condition is not satisfied~\cite[Sec.~B3]{foot:2}.
In principle, the minimization procedure also outputs the force vector $\vec f$, but extracting small forces from it requires an 
even heavier use of quad precision arithmetics than what we have used for the energy minimization~\cite[Sec.~B3]{foot:2}.
We instead determine $\vec f$ as the zero mode of $\NN$ for this packing~\cite[Sec.~A6]{foot:2}, for which double precision arithmetics suffices.
Because $\NN$ is sparse, the eigenvector corresponding to the zero-mode can efficiently be extracted with the Lanczos algorithm~\cite[Sec.B3]{foot:2}, as implemented in Mathematica~\cite{La50,Mathematica10}.

\section{Results}
Figure~\ref{fig:forces} gives the cumulative 
force distribution $G(f)=\int_0^f P(f')df'$ for $d$=2, 3, and 4. In all cases, a power-law scaling at weak forces is detected, but the value of $\theta_\mathrm{f}$ is found to increase with $d$. Recall that over the same $d$ range $\gamma$ remains robustly constant~\cite{CCPZ12} 
and is consistent with 
$\g=0.41269$ from the $d=\infty$ solution~\cite{CKPUZ14,CKPUZ14b}. 
The changing value of $\theta_{\rm f}$ with $d$ is therefore inconsistent with the scaling relations between exponents $\theta$ and $\gamma$ 
in a marginally stable phase~\cite{Wy12,DeGiuli14,DLBW14,LDW13}.

\begin{figure}
\includegraphics[width=0.8\columnwidth]{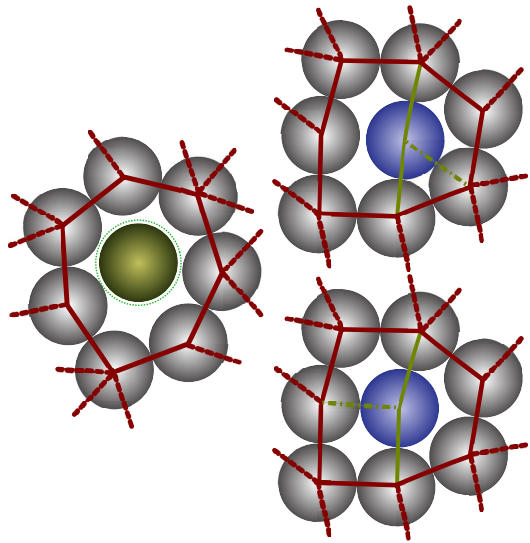}
\caption{Schematic depiction of (left) rattlers and (right) bucklers. At jamming, rattlers (also known as floaters -- green) are not part of the force network. Their neighbors (grey) are part of the force network (red lines) and form a rigid cage within which the rattler can freely move (dashed green line). By contrast, bucklers are part of the force network (red and green lines), but breaking their weakest contact (dot-dashed green line) only creates a localized excitation (top and bottom). Bucklers are thus typically particles with $d$ nearly co-planar stronger contacts and a weaker $d+1$th contact that balances the resulting normal force.
}
\label{fig:sketch}
\end{figure}

\begin{figure}
\includegraphics[width=\columnwidth]{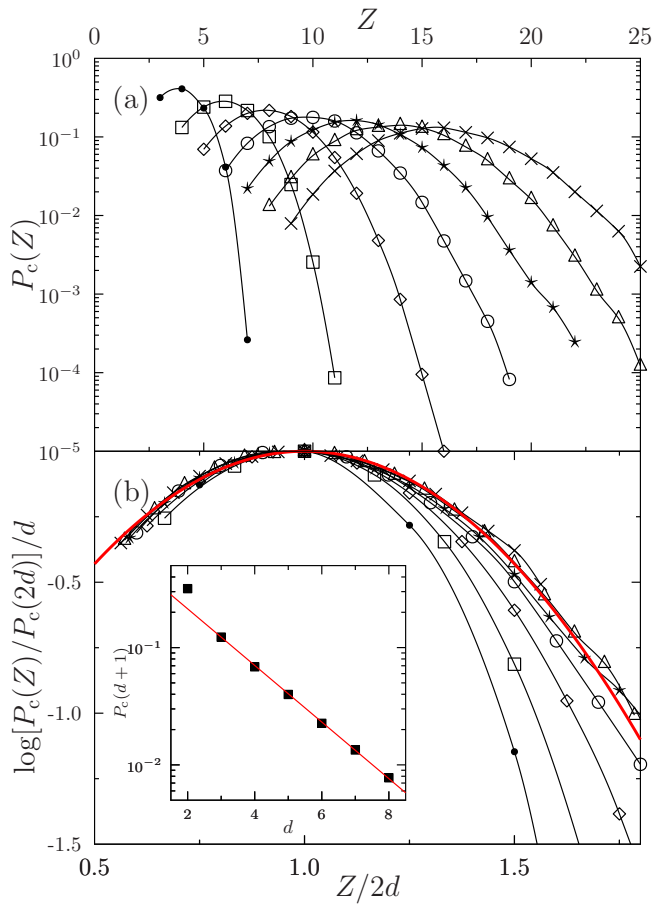}
\caption{(a) Probability distribution of the number of contacts $Z$ for particles within the force network (neglecting rattlers) at jamming in $d=2\ldots8$, from left to right, for the protocol described in the text. Note that the distribution peaks around $Z=2d$. (b) Rescaling these distributions using a large-deviation form shows the results to converge fairly quickly with $d$. For this preparation protocol, the form is nearly Gaussian (red line). The large deviation form suggests that the proportion of particles with $Z = d+1$  (and thus of bucklers) decays exponentially with $d$, (inset) as is explicitly observed in finite $d$ (red line).}
\label{fig:Zdist}
\end{figure}

Soft mode excitations suggest a possible way to resolve this paradox~\cite{LDW13,DeGiuli14,DLBW14}. 
The proposed mechanism for localized excitations is for a particle to have all but one of its contact vectors be nearly co-planar~\cite{LDW13}. The remaining contact must then necessarily be weak (by force balance). Breaking that contact should then result in facile back and forth buckling. Because this motion does not affect the rest of the packing much, the resulting excitation is fairly localized (Fig.~\ref{fig:sketch}). Although a nearly coplanar arrangement of neighbors is formally possible for a particle  with any $Z$, in a sufficiently disordered (non-crystalline) system it grows increasingly unlikely with $Z$. This arrangement is therefore most likely to occur for particles that have the minimal $Z$ for maintaining local stability, i.e., $Z=d+1$. 
Particles with $Z=d+1$ contacts and for which one contact is weak (we dub them {\it bucklers}) 
are also overwhelmingly likely to have its other $d$ contact particles be nearly coplanar with its center of mass~\cite[Sec.~B4]{foot:2}. Any other arrangement would entail the presence of at least two weak contacts, which is highly unlikely. In summary, with high probability, all bucklers have $d+1$ contacts and all particles with $d+1$ contacts and a weak force are bucklers. 
In Fig.~\ref{fig:forces}, we consider $P_{\ell}(f)$ the distribution of all forces involving particles with $d+1$ contacts (and thus all bucklers), 
and $P_{\rm e}(f)$ that of the remaining contact forces.
This breakdown cleanly separates the power-law regimes for $\theta_{\mathrm{e}}$ and $\theta_{\ell}$. Remarkably (this is our main result), $P_{\ell}(f) \sim f^{\theta_\ell}$ while $P_{\rm e}(f) \sim f^{\theta_{\rm e}}$, with exponents
independent of $d$ and consistent with the $d=\io$ solution and the scaling relations.

This finding also provides an explanation for the behavior of $\theta_{\rm f}$. In order to see why, let us define
the distribution of the number of contacts $P_{\rm c}(Z)$ ($\sum_Z Z P_{\rm c}(Z) \approx 2 d$).
The fraction of forces adjacent to bucklers is then $n_{\ell} = (d+1) P_{\rm c}(d+1)/2d$, and
the total force distribution
\beq
P(f) = n_{\ell} P_{\ell}(f) + (1-n_{\ell}) P_{\rm e}(f) \ .
\eeq
For large $d$, it is reasonable to expect that $P_{\rm c}(Z) \sim e^{d \h_{\rm c}(Z/d)}$ 
becomes strongly peaked (in relative terms) around the average of $Z$,
and is roughly Gaussian around that average. Figure~\ref{fig:Zdist} confirms this hypothesis, and as a result
$P_{\rm c}(d+1)$ and $n_{\ell} \sim P_{\rm c}(d+1)/2$ both decrease exponentially with $d$. 
For small $f$, it follows that $
P(f) \sim n_{\ell} f^{\theta_\ell} + (1-n_{\ell}) f^{\theta_{\rm e}}$.
Hence, it is correct that asymptotically one should observe $\theta_\mathrm{f} =\mathrm{min}(\theta_\mathrm{e}, \theta_\ell)$,
but only when $n_{\ell} f^{\theta_\ell} \gg f^{\theta_{\rm e}}$, i.e., for forces exponentially small in $d$.
This result explains why no trace of bucklers nor of localized modes can be found in the 
$d=\infty$ solution, for which $n_{\ell}=0$ and thus $\theta_{\rm f} = \theta_{\rm e}$.
It also suggests that the contribution of bucklers cannot be perturbatively detected around that solution either.
In the force regime that is numerically (and experimentally) accessible in low $d$, 
an effective mixed value for $\theta_\mathrm{f}$ is observed, which is close to $\theta_\ell$
in $d=2$ and increases smoothly towards $\theta_{\rm e}$ as $d$ increases, reflecting the 
systematic decrease of $n_\ell$.

\section{Conclusion} Our results demonstrate that the jamming criticality remains robustly constant for $d\geq2$, although the spurious contribution of rattlers and bucklers must be excluded from the structural analysis in order to cleanly detect it. This remarkable outcome confirms that certain aspects of 
mean-field marginality subsist in finite-dimensional systems, including in experimentally-relevant $d=2$ and 3~\cite{CKPUZ14,CKPUZ14b}. These results should therefore be experimentally verifiable.

The theoretical explanation as to why long-wavelength fluctuations do not renormalize the properties of jamming criticality in these systems remains thus far unanswered (see~\cite{UB14}
for a preliminary investigation). 
One may argue that the complete absence of thermal fluctuations at jamming, and/or the presence of long-ranged elastic interactions~\cite{MW14},
may play a role. 
This observation would then suggest that marginal systems with other types of disorder, be it related to constraint satisfaction 
or size dispersity, may also exhibit similarly robust mean-field criticality upon approaching their ground state.
It is important to note, however, that this universality does not imply that away from jamming thermal fluctuations may not 
destroy the mean-field marginal state structure and round off the associated phase transitions~\cite{UB14}.

\begin{acknowledgments}
We acknowledge many fruitful exchanges with L.~Berthier, G.~Biroli, E.~DeGiuli, C.~Goodrich, Y.~Jin, E.~Lerner, A.~Liu, P.~Urbani, and M.~Wyart. 
We gratefully acknowledge help from C. Barbieri at Sapienza and M.~Peterson at Duke with computational assistance. 
The European Research Council has provided financial support through ERC grant agreement no.~247328. EIC thanks the NSF for support under CAREER Award No. DMR-1255370. PC acknowledges support from the National Science Foundation Grant no. NSF DMR-1055586 and the Sloan Foundation.  The use of the ACISS supercomputer is supported under a Major Research
Instrumentation Grant, Office of Cyber Infrastructure, No. OCI-0960354.
\end{acknowledgments}

\appendix
\begin{widetext}

\section{Isostaticity, force network, and soft modes}
\label{sec:Theory}
This section consolidates and completes the results of Refs.~\cite{LDW13,DLBW14,LDDW14} that are needed for the presentation of the main text.
Note that in order to highlight the logical flow of the discussion, we denote the main results and summaries by (R1), (R2), etc.

Consider a packing with $k = 1 \cdots N$ particles in $\alpha =1 \cdots d$ dimensions.
Two particles are in contact if $|\textbf{r}_i - \textbf{r}_j| = \sigma_{ij}$, where $\sigma_{ij}=(\sigma_i+\sigma_j)/2$ is the sum of the particle radii.
We define $\partial k$ the set of particles that are in contact with particle $k$.
A contact is an ordered pair $\la ij \ra$ with $i<j$ that we consider as a single index $\la ij \ra = 1 \cdots N_{\mathrm{c}}$, where $N_{\rm c}$ is
the total number of contacts.
On each contact there is a scalar contact force $f_{ij} = f_{ji}$, 
and we define $\vec f = \{ f_{ij} \}$, a vector that lives in a $N_{\mathrm{c}}$-dimensional vector space.
The particles positions are $r_{k\alpha}$ and the external forces are $F_{k\alpha}$. We use bold letters, 
e.g., $\mathbf{r}_k$ and $\mathbf{F}_k$, to denote $d$-dimensional vectors.
By contrast, 
we define $\vec F = \{ F_{k\alpha} \}$ and $\vec r = \{ r_{k\alpha} \}$ as vectors that
live in the $Nd$-dimensional vector space.
We also define the contact vector $\mathbf{n}_{ij} = (\mathbf{r}_j - \mathbf{r}_i)/|\mathbf{r}_j - \mathbf{r}_i|$. Note that $\mathbf{n}_{ij} = -\mathbf{n}_{ji}$.

\subsection{Force balance equations}
\label{sec:fb}

The force balance equations 
\beq\label{eq:forcebalance}
F_{k\alpha}  = \sum_{j \in \partial k} n_{jk}^\alpha f_{jk} 
\eeq
can be written in matrix notation as
\beq\label{forcebalanceS}
F_{k\alpha}  = \sum_{\la i j \ra} (\SS^T)^{\la ij \ra}_{k \alpha} f_{ij} =  \sum_{\la i j \ra} \SS_{\la ij \ra}^{k \alpha} f_{ij} 
\ \ \ \ \ \
\Rightarrow
\ \ \ \ \ \ 
\vec F = \SS^T \vec f \ ,
\eeq
where $\SS$ is a $N_{\mathrm{c}} \times Nd$ matrix with elements
\beq\label{SSdef}
\SS_{\la ij \ra}^{k \alpha} = ( \d_{jk} - \d_{ik}) n_{ij}^\alpha \ .
\eeq
We also define a $N_{\mathrm{c}} \times N_{\mathrm{c}}$ symmetric matrix $\NN = \SS \, \SS^T$ with elements
\beq\label{eq:M}
\NN_{\la ij \ra}^{\la lm \ra} = \sum_{k\alpha} \SS_{\la ij \ra}^{k \alpha}\SS_{\la lm \ra}^{k \alpha}
= (\d_{il} - \d_{mi} - \d_{lj} + \d_{mj} ) \,  \textbf{n}_{ij} \cdot \textbf{n}_{lm} \ ,
\eeq
and keep in mind that the zero modes of $\SS^T$ are also zero modes of $\NN$, but that $\NN$ can have additional zero modes.

In the following, we consider two possible situations.
\begin{enumerate}
\item Imposing $F_{k\alpha}=0$, which corresponds to a mechanical equilibrium in absence of external forces, such as under periodic boundary conditions.
Then Eqs.~\eqref{eq:forcebalance} 
are $Nd$ {\it homogeneous} linear equations for the $N_{\mathrm{c}}$ contact forces. 
Note, however, that $\sum_{k=1}^N F_{k\alpha}
=  \sum_{k, j \in \partial k} n_{jk}^\alpha f_{jk} =0$ 
because $f_{ij}$ is symmetric
while $\textbf{n}_{ij}$ is antisymmetric. This condition expresses the global translational invariance of the system. 
As a consequence, we get that $d$ equations are linearly dependent of the others, and thus only $(N-1)d$ equations are independent. 
\beq\tag{R1}
\boxed{
\begin{aligned}
&\text{In absence of external forces, Eqs.~\eqref{eq:forcebalance} admit $\max[N_{\mathrm{c}} - (N-1)d,0]$} \\
&\text{non-zero linearly independent solutions.}
\end{aligned} 
}
\eeq

\item Imposing mechanical equilibrium under non-zero external forces that satisfy $\sum_k F_{k\alpha} =0$. This situation corresponds, for instance, to the presence of external confining walls that fix the center of mass of the packing. In this case,
Eqs.~\eqref{eq:forcebalance} are {\it inhomogeneous}. 
\beq
\tag{R2}
\boxed{
\begin{aligned}
&\text{In presence of external forces, Eqs.~\eqref{eq:forcebalance} admit a unique solution for $N_{\mathrm{c}} = (N-1)d$, 
} \\
&\text{while 
for $N_{\mathrm{c}} > (N-1) d$ the solutions form a linear space of dimension $N_{\mathrm{c}} - (N-1)d$. }
\end{aligned}
}
\eeq

\end{enumerate}

\subsection{Particle displacements}
\label{sec:H}

For a given packing, we now consider a displacement $\d r_{i\alpha}$ of particle $i$ in direction $\alpha$ and 
the distance between a pair of particles $\rho_{ij} = | \mathbf{r}_i - \mathbf{r}_j |$. 
To linear order,
\beq\label{rij}
\d \rho_{ij} = \sum_{\alpha} n_{ij}^\alpha (\d r_{j\alpha} - \d r_{i\alpha}) \ ,
\ \ \ \ \ \
\Rightarrow
\ \ \ \ \ \ 
\d \vec \rho = \SS \d \vec r \ ,
\eeq
where $\d \vec \rho = \{ \d \rho_{ij} \}$ lives in the $N_{\mathrm{c}}$-dimensional vector space, while $\d \vec r = \{ \d r_{k\alpha} \}$ lives
in the $Nd$-dimensional vector space. Displacements $\d \vec r$ that leave the distances in the packing invariant
should satisfy $0 = \SS \d \vec r$. Equations~\eqref{rij} thus always admit $d$ trivial solutions $\d r_{k \alpha} = \d_{\alpha,\alpha'}$
that correspond to uniform translations of the packing in the $d$ available directions. 

Now, consider soft harmonic spheres that are almost at jamming. The potential energy is 
\begin{equation}
U = \kappa \sum_{\la ij \ra} ( |\textbf{r}_i - \textbf{r}_j| - \sigma_{ij} )^2\theta(\sigma_{ij}-|\textbf{r}_i - \textbf{r}_j|),
\label{eq:harmonic}
\end{equation}
where $\theta(r)$ is the Heaviside function, and the stiffness $\kappa$ is set to unity without loss of generality. Thus, $|\textbf{r}_i - \textbf{r}_j| = \sigma_{ij} - \e_{ij}$ for contacts, and $|\textbf{r}_i - \textbf{r}_j| > \sigma_{ij}$ for non-contacts. 
{\it Assuming that contacts cannot be opened}, 
in the limit $\e \to 0$ the $Nd\times Nd$ elements of the Hessian matrix are 
\beq
\HH_{i\alpha}^{j\alpha'} = 
\frac{\partial^2 U}{\partial r_{i\alpha} \partial r_{j\alpha'}} = \d_{ij}  \sum_{k\in \partial i} n_{ki}^\alpha n_{ki}^{\alpha'} - n_{ij}^\alpha n_{ij}^{\alpha'} \d(\la ij \ra) \ ,
\eeq
where $\d(\la ij \ra) = 1$ if $ij$ are in contact and zero otherwise. It follows that $\HH = \SS^T  \SS$, which can be shown using Eq.~\eqref{SSdef}:
\beq\label{Hdef}
\begin{split}
( \SS^T \SS )_{i\alpha}^{j\alpha'} &= \sum_{\la kl \ra} \SS_{\la kl \ra}^{i \alpha}   \SS_{\la kl \ra}^{j \alpha'}
= \sum_{\la kl \ra} ( \d_{li} - \d_{ki}) n_{kl}^\alpha ( \d_{lj} - \d_{kj}) n_{kl}^{\alpha'}
= \sum_{\la kl \ra}  n_{kl}^\alpha n_{kl}^{\alpha'} ( \d_{li} \d_{lj} - \d_{ki} \d_{lj} - \d_{li} \d_{kj} + \d_{ki} \d_{kj}) \\ 
&= \d_{ij}  \sum_{k\in \partial i} n_{ki}^\alpha n_{ki}^{\alpha'} 
- n_{ij}^{\alpha'} n_{ij}^{\alpha'} \d(\la ij \ra) = \HH_{i\alpha}^{j\alpha'} \ .
\end{split}\eeq
\beq
\tag{R3}
\boxed{
\begin{aligned}
&\text{The energy of small displacements, i.e., displacements small enough that they do not open any contact, is} \\
&U \sim \d \vec r^T \HH \d \vec r = \d \vec r^T \SS^T \SS \d \vec r = (\SS \d\vec r)^2 \ .
\end{aligned}
}
\eeq

\subsection{Eigenvalue Algebra}
\label{algebra}

From the definition of the $N_{\mathrm{c}} \times Nd$ matrix $\SS$ in Eq.~\eqref{SSdef}, of the $N_{\mathrm{c}} \times N_{\mathrm{c}}$ symmetric matrix $\NN = \SS \, \SS^T$ in Eq.~\eqref{eq:M},
and the $Nd \times Nd$ symmetric matrix $\HH = \SS^T  \SS$ in Eq.~\eqref{Hdef}, it follows that, for all $p$, the eigenvalues $\lambda$ obey
\beq
\sum_{i=1}^{Nd} \l_{\HH,i}^p =
\Tr \HH^p = \Tr \NN^p = \sum_{i=1}^{N_{\mathrm{c}}} \l_{\NN,i}^p 
 \ ,
\hskip30pt
\forall p \ .
\eeq
This result implies that the non-zero eigenvalues of $\NN$ and $\HH$ are identical.

Recall that $\HH$ always has $d$ zero eigenvalues due to translational invariance.
Then, if $N_{\mathrm{c}} \geq (N-1) d$, the matrix $\NN$ must have $z_\NN = N_{\mathrm{c}} - N d + d = N_{\mathrm{c}} - (N-1)d$ zero eigenvalues.
Similarly, if $N_{\mathrm{c}} \leq (N-1) d$, the matrix $\HH$ must have $z_\HH = N d - N_{\mathrm{c}}$ zero eigenvalues. 
\beq\tag{R4}
\boxed{
\begin{aligned}
\text{The number} &\text{ of zero modes of the matrices $\NN$ and $\HH$ is} \\
&z_\NN  = \begin{cases}
N_{\mathrm{c}} - (N-1)d & \text{if } N_{\mathrm{c}} \geq (N-1) d \\
0 & \text{if } N_{\mathrm{c}} \leq (N-1) d 
\end{cases} \ , \\
&z_\HH  = d+ \begin{cases}
0 & \text{if } N_{\mathrm{c}} \geq (N-1) d \\
(N-1) d - N_{\mathrm{c}} & \text{if } N_{\mathrm{c}} \leq (N-1) d 
\end{cases} \ .
\end{aligned}
}
\eeq

\subsection{Floppy modes}
\label{floppy}

Now, consider a packing with $N_{\mathrm{c}}$ contacts. We select one of these contacts, $\la ij\ra$, which
for notational simplicity we label $\tau = \la ij \ra$ (for two particles touching).
We want to displace the particles by $\d\vec r^{(\tau)}$,
such that all distances $\r_{kl}$ in contacts $\la kl \ra \neq \tau$ remain unchanged, while the contact $\tau$ is opened by an infinitesimal amount.
The resulting excitation is the {\it floppy mode} associated with opening contact $\tau$. It is floppy, because the contact is opened, hence it does not contribute anymore to the system energy,
and all the other contacts $\langle kl\rangle$ remain at a distance $\sigma_{kl}$, hence they also do not contribute to the energy. The total system energy thus remains zero.
To lowest order, the variation of the distance $\rho_{kl}$ is given by Eq.~\eqref{rij}, and we want to impose
 $\d \rho_{ij} =1$ (or any other infinitesimal amount, because the equations are in any case linear), while
$\d \r_{kl} =0$ for all other contacts.
We thus want to solve
\beq\label{eqa}
\d \rho_{kl}^{(\tau)} = \textbf{n}_{kl} \cdot (\d \textbf{r}_k^{(\tau)} - \d \textbf{r}_l^{(\tau)} ) = \d_{\tau,\la k l \ra} 
\ \ \ \ \ \
\Rightarrow
\ \ \ \ \ \ 
\d \vec \rho^{(\tau)} = \SS \d \vec r^{(\tau)} = \vec \tau \ ,
\eeq
where the $N_{\mathrm{c}}$-dimensional vector $\vec \tau$ belongs to the space of contacts with components $\tau_{\la kl \ra} = \d_{\tau,\la k l \ra}$,
i.e. it is equal to 1 for contact $\tau$ and zero for all other contacts.
Because we want to exclude global translations of the packing from the solutions of Eq.~\eqref{eqa}, we impose
$\sum_{k=1}^N \d \mathbf{r}^{(\tau)}_k =0$. In this way, {\it (i)} the vector $\d\vec r^{(\tau)}$ is orthogonal to the $d$ zero modes of $\SS$,
and {\it (ii)} it is parametrized by $(N-1)d$ independent variables.
Equation~\eqref{eqa} is thus a set of $N_{\mathrm{c}}$ {\it non-homogeneous} linear equations for $(N-1)d$ independent variables.
\beq
\tag{R5}
\boxed{
\begin{aligned}
&\text{Equations~\eqref{eqa} admit a unique solution for $N_{\mathrm{c}} = (N-1)d$, 
} \\
&\text{while 
for $N_{\mathrm{c}} < (N-1) d$ the solutions form a linear space of dimension $ (N-1)d - N_{\mathrm{c}}$. }
\end{aligned}
}
\eeq

\subsection{Response to a dipolar force field}

By applying $\SS^T$ to Eq.~\eqref{eqa} we obtain
\beq\label{eqb}
\SS^T \SS \d \vec r^{(\tau)} = \HH \d \vec r^{(\tau)} = \SS^T \vec \tau \ .
\eeq
From Sec.~\ref{sec:fb}, we know that matrix $\SS^T$ has $N_{\mathrm{c}} - (N-1)d$ zero modes, and therefore three scenarios are possible.
\begin{enumerate}
\item If $N_{\mathrm{c}} < (N-1)d$, then Eq.~\eqref{eqa} has many solutions, and $\SS^T$ has no zero modes. \\
\beq
\tag{R6}
\boxed{
\begin{aligned}
&\text{If $N_{\mathrm{c}} < (N-1)d$, Eq.~\eqref{eqa} and Eq.~\eqref{eqb} have the same solutions.} \\
\end{aligned}
}
\eeq

\item If $N_{\mathrm{c}} > (N-1)d$, then Eq.~\eqref{eqa} has no solutions. Hence, $\SS^T$ has some zero modes, and Eq.~\eqref{eqb} can admit solutions if 
\beq\label{eqapp1}
\SS \d \vec r^{(\tau)} - \vec \tau = \vec f^{(\tau)} \ ,
\eeq
where $\vec f^{(\tau)}$ is one of the zero modes of $\SS^T$. In general we do not know how many solutions of Eq.~\eqref{eqapp1} exist.
However, the vector $ (\SS^T \vec \tau)_{ k\alpha } = \SS_{\tau}^{k \alpha} = ( \d_{jk} - \d_{ik}) n_{ij}^\alpha$ is clearly orthogonal to the $d$ trivial zero modes of $\SS$ and $\HH$,
because $\sum_k  (\SS^T \vec \tau)_{ k\alpha } = \sum_k ( \d_{jk} - \d_{ik}) n_{ij}^\alpha =0$. 
If $N_{\mathrm{c}} \geq (N-1)d$, these are the only zero modes of $\HH$, and therefore we can invert $\HH$ by restricting ourselves to the space orthogonal to the $d$ trivial
zero modes. We then have
\beq\label{eqc}
\d \vec r^{(\tau)} = \HH^{-1} \SS^T \vec \tau \ .
\eeq
\beq
\tag{R7}
\boxed{
\begin{aligned}
&\text{If $N_{\mathrm{c}} > (N-1)d$, Eq.~\eqref{eqa} has no solutions and Eq.~\eqref{eqb} has a unique solution given by Eq.~\eqref{eqc}.} \\
\end{aligned}
}
\eeq

\item
If $N_{\mathrm{c}} = (N-1)d$, then Eq.~\eqref{eqa} has a unique solution and $\SS^T$ has no zero modes. Hence, Eq.~\eqref{eqb} also has a unique solution. 
In addition, $\HH$ only has the trivial zero modes
so the reasoning from the previous point applies. 
\beq
\tag{R8}
\boxed{
\begin{aligned}
&\text{If $N_{\mathrm{c}} = (N-1)d$, Eq.~\eqref{eqc} is the unique solution of both Eq.~\eqref{eqa} and Eq.~\eqref{eqb}.} \\
\end{aligned}
}
\eeq

\end{enumerate}
Note that Eq.~\eqref{eqc} can be interpreted as the response to a dipolar force field. Suppose that we take a packing in equilibrium with 
zero external forces and we apply an external force $\ee \SS^T \vec \tau$ to it. Recall that $ (\SS^T \vec \tau)_{ k\alpha } = ( \d_{jk} - \d_{ik}) n_{ij}^\alpha$
hence we are applying a force $\mathbf{F}_i = \ee \mathbf{n}_{ij}$ on particle $i$ and $\mathbf{F}_j = -\ee \mathbf{n}_{ij}$ on particle $j$, i.e., a dipolar force. 
For small $\ee$, minimizing
the energy gives
\beq
\frac{\partial U}{\partial r_{i\alpha}}(\vec r + \ee \d\vec r) =\ee (\SS^T \vec \tau)_{ i\alpha }
\ \ \ \ \ \
\Rightarrow
\ \ \ \ \ \ 
\sum_{j\alpha'} \HH_{i\alpha}^{j\alpha'} \d r_{j\alpha'} = (\SS^T \vec \tau)_{ i\alpha } \ ,
\eeq
which coincides with Eq.~\eqref{eqb} and is solved by Eq.~\eqref{eqc} for $N_{\mathrm{c}} \geq (N-1)d$.

\subsection{Isostaticity in absence of external forces}

We consider the special case $N_{\mathrm{c}} = (N-1)d + 1$ in absence of external forces, which corresponds to a packing under periodic boundary conditions. 
In this case,
we have that:
\begin{itemize}
\item $z_\NN=1$, hence $\NN$ has a unique zero mode (Sec.~\ref{algebra});
\item the force balance equation $\SS^T \vec f=0$ has a unique solution (Sec.~\ref{sec:fb});
\item because $\NN = \SS \SS^T$, the solution of $\SS^T \vec f$ must be the unique zero mode of $\NN$.
\end{itemize}
Hence, the contact forces $\vec f$ are given by the unique zero mode of $\NN$ and are fixed up to an overall scale factor (the global pressure), which is left free because there are no external forces. We also have $z_\HH=d$ (Sec.~\ref{algebra}), hence the only zero modes of the small displacement matrix are those corresponding to
global translations of the packing, and there are no floppy modes (Sec.~\ref{floppy}).

\beq
\tag{R9}
\boxed{
\begin{aligned}
&\text{Under periodic boundary conditions, isostaticity corresponds to $N_{\mathrm{c}} = (N-1)d + 1$, and:} \\
&\text{the forces are determined by the packing geometry through $\NN \vec f =0$, up to an overall scale factor;} \\
&\text{the packing is mechanically stable in the sense that $\HH$ has no non-trivial zero (or floppy) modes;} \\
&\text{the response to a dipolar force is given by Eq.~\eqref{eqc}.} \\
\end{aligned}
}
\eeq

\subsection{Isostaticity in presence of external forces}

We consider the special case $N_{\mathrm{c}} = (N-1)d$ in presence of external forces, which corresponds to a packing confined by walls. In this case, we have that:
\begin{itemize}
\item the force balance equation has a unique solution (Sec.~\ref{sec:fb});
\item contact forces are fully determined by the external forces;
\item $\HH$ has no zero modes apart from the trivial ones. 
\end{itemize}
Hence, small fluctuations that do not break contacts are stable. However, each contact $\tau$ corresponds to a unique floppy mode, given by Eq.~\eqref{eqc}, that breaks contact $\tau$ keeping all the other
contact distances fixed. This \emph{non-linear} soft mode has non-zero energy only because of the external forces.
These soft modes are the ones used in the stability analysis of Ref.~\cite{LDW13}.
\beq
\tag{R10}
\boxed{
\begin{aligned}
&\text{In presence of external walls, isostaticity corresponds to $N_{\mathrm{c}} = (N-1)d$, and:} \\
&\text{the forces are uniquely determined by the external forces;} \\
&\text{the packing is mechanically stable in the sense that $\HH$ has no zero modes;} \\
&\text{floppy modes allow a contact to open without affecting the other contacts;} \\
&\text{they are given
by Eq.~\eqref{eqc}, and their energy depends only on the external confining forces.} \\
\end{aligned}
}
\eeq

\subsection{Other harmonic systems (including hard spheres)}

We now consider the generalization of the discussion of Sec.~\ref{sec:H}
to packings of other types of harmonic spheres near jamming. 
First, we consider a potential energy $U = \sum_{\la ij \ra} \k_{ij} ( |\mathbf{r}_i - \mathbf{r}_j| - \sigma_{ij} )^2 \theta(\sigma_{ij}-|\textbf{r}_i - \textbf{r}_j|)$ with heterogeneous
stiffnesses $\k_{ij}$.
Assuming that contacts cannot be opened, the elements of the Hessian matrix are then
\beq\label{eq:Hp}
\HH_{i\m}^{j\alpha'} = 
\frac{\partial^2 U}{\partial r_{i\alpha} \partial r_{j\alpha'}} = \d_{ij}  \sum_{k\in \partial i} \k_{ki} n_{ki}^\alpha n_{ki}^{\alpha'} - \k_{ij} n_{ij}^\alpha n_{ij}^{\alpha'} \d(\la ij \ra) \ ,
\eeq
Second, we consider a system with a potential energy $U = \kappa \sum_{\la ij \ra} \big| |\mathbf{r}_i - \mathbf{r}_j| - \sigma_{ij} \big|^\nu \theta(\sigma_{ij}-|\textbf{r}_i - \textbf{r}_j|)$. The force on contact $\langle ij\rangle$ is then
\beq
f_{ij} \propto \big| |\mathbf{r}_i - \mathbf{r}_j| - \sigma_{ij} \big|^{\nu-1} \ ,
\eeq
and the effective stiffness is 
\beq\label{eq:kf}
\k_{ij} \propto \big| |\mathbf{r}_i - \mathbf{r}_j| - \sigma_{ij} \big|^{\nu-2} \propto f_{ij}^{(\nu-2)/(\nu-1)} \ .
\eeq
At a isostatic point under periodic boundary conditions, contact forces are uniquely determined by the force balance equations, and are thus independent of the particular choice of
potential. Once the forces are determined, the effective stiffnesses can be obtained via Eq.~\eqref{eq:kf}. Plugging this result in Eq.~\eqref{eq:Hp},
we obtain a matrix $\HH$ that gives the small fluctuations associated with this modified potential, provided the harmonic approximation holds.

A special case of interest is that of hard spheres, which corresponds to $\nu \to 0$ in the absence of thermal excitations, and thus $\k_{ij} \propto f_{ij}^2$. The corresponding Hessian matrix has elements
\beq
\HH_{i\alpha}^{j\alpha'} = 
\frac{\partial^2 U}{\partial r_{i\alpha} \partial r_{j\alpha'}} = \d_{ij}  \sum_{k\in \partial i} f_{ki}^2 n_{ki}^\alpha n_{ki}^{\alpha'} - f_{ij}^2 n_{ij}^{\alpha} n_{ij}^{\alpha'} \d(\la ij \ra) \ .
\eeq
In this case, we can define a modified matrix $\tilde{\SS}$ with elements
\beq\label{SSpdef}
\tilde{\SS}_{\la ij \ra}^{k \alpha} = f_{ij} ( \d_{jk} - \d_{ik}) n_{ij}^\alpha \ .
\eeq
Using Eq.~\eqref{SSpdef}, one can then show that $\HH = \tilde{\SS}^T  \tilde{\SS}$ by noting that
\beq\begin{split}
( \tilde{\SS}^T \tilde{\SS} )_{i\alpha}^{j\alpha'} &= \sum_{\la kl \ra} \tilde{\SS}_{\la kl \ra}^{i \alpha}   \tilde{\SS}_{\la kl \ra}^{j \alpha'}
= \sum_{\la kl \ra} f_{kl}^2 ( \d_{li} - \d_{ki}) n_{kl}^\alpha ( \d_{lj} - \d_{kj}) n_{kl}^{\alpha'}
\\ 
&= \d_{ij}  \sum_{k\in \partial i} f_{ki}^2 n_{ki}^\alpha n_{ki}^{\alpha'} 
- f_{ij}^2 n_{ij}^{\alpha} n_{ij}^{\alpha'} \d(\la ij \ra) = \HH_{i\alpha}^{j\alpha'} \ .
\end{split}\eeq
Note also that because the contact forces are determined by the force balance condition in Eq.~\eqref{forcebalanceS}, the modified matrix
$\tilde{\SS}$ must have a zero eigenvector with constant components.

\section{Numerical simulations}
This section completes the numerical details used in order to obtain isostatic configurations and to extract their force network.

\subsection{Isostaticity Considerations}
The exact solution for $d=\infty$ finds that $\lim_{N\to\io} (2N_{\mathrm{c}}/N) = 2d$ for jammed packings, independently of the jamming density, but does not explicitly provide the finite $N$ corrections. From that viewpoint, any solution that
gives $N_{\mathrm{c}} = d N + \OO(1)$ may thus be acceptable. One might nevertheless wonder what $N_{\mathrm{c}}$ should be, for a finite $N$ systems, in order to most efficiently converge to the thermodynamic limit. Is it sufficient to have
$N_{\mathrm{c}} = d N+ \OO(1)$, or does one need exactly $N_{\mathrm{c}} = (N-1)d +1 \equiv N_{\mathrm{c}}^{\mathrm{iso}}$ (under cubic periodic boundary conditions; other choices of boundary conditions have different $\OO(1)$ corrections~\cite{HST13})?
Early numerical simulations~\cite{OLLN02,DTS05}, including our own~\cite{CCPZ12}, 
did not pay much attention to this issue. The protocols used did not exactly result in $N_{\mathrm{c}} = N_{\mathrm{c}}^{\mathrm{iso}}$, typically because of 
compression rates that were too rapid, incorrect stopping criteria, insufficient numerical precision, etc. Although the packings had $N_{\mathrm{c}} - N_{\mathrm{c}}^{\mathrm{iso}}\neq 0$, most of the jamming phenomenology was nevertheless
found to be robustly conserved from one set of simulations to another.
Recently, the results of several more careful simulation protocols~\cite{GLN12,HST13,LDW13} have, however, highlighted the importance
of having exactly $N_{\mathrm{c}} = N_{\mathrm{c}}^{\mathrm{iso}}$ to observe some key aspects of jamming criticality in finite systems. In particular, the theoretical
analysis of Ref.~\cite{Wy12} relies heavily on packings being strictly isostatic with $N_{\mathrm{c}} = N_{\mathrm{c}}^{\mathrm{iso}}$. For this reason, in this study we exclusively consider packings with $N_{\mathrm{c}} = N_{\mathrm{c}}^{\mathrm{iso}}$, which allows us
to directly apply the analysis outlined in Sec.~\ref{sec:Theory}. It may nonetheless be interesting to check to what extent measurements of the critical exponents in finite $N$ are affected by $N_{\mathrm{c}} \approx  N_{\mathrm{c}}^{\mathrm{iso}}$, but this analysis is left for future work.

\subsection{Detailed Numerical Minimization Protocol}
The key difficulty in obtaining packings with  $N_{\mathrm{c}}^{\mathrm{iso}}$ is distinguishing between contacts and near contacts. In finite-precision arithmetics, this challenge follows from the gap distribution for near contacts being singular and the contact force distribution having a fat power-law tail at weak forces. Hence, being insufficiently close to jamming results in ambiguities in the force network determination. Compounding this difficulty is the need to precisely remove rattlers, as they are not part of the force network itself. In order to produce a packing that is truly isostatic, one thus need to impose that the distance between the contact spheres should be very near the particle diameter, with a precision that increases with $N_{\mathrm{i}}$. For instance, in a system with $N_{\mathrm{i}}=16384$ in $d=4$, with high probability at least one near contact is only $\OO(10^{-9})\sigma_{ij}$ away from contact. \emph{All contact pairs} must therefore be known to within a higher precision than that value.

\begin{figure}
\includegraphics[width=0.6\columnwidth]{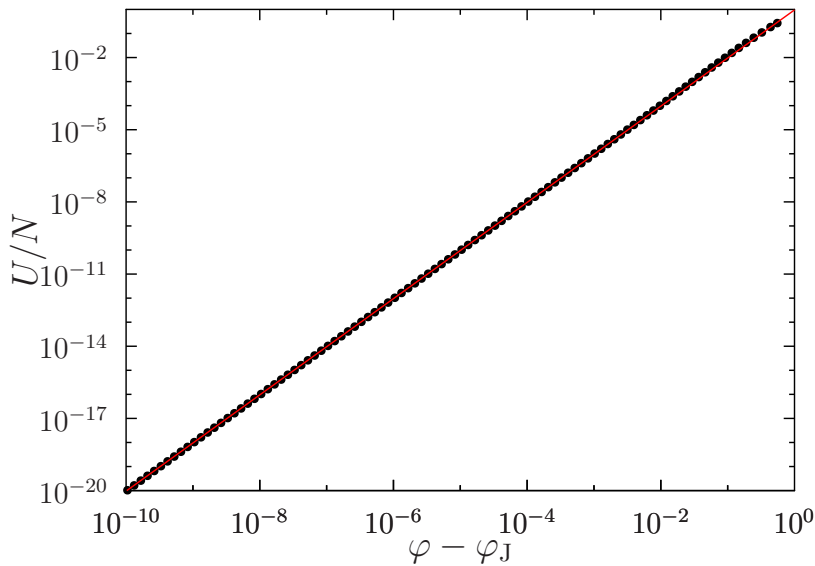}
\caption{Energy per particle $U/N\propto(\varphi - \varphi_{\mathrm{J}})^2$ for the successive minimization of a representative system with $N_{\mathrm{i}}=16384$ in $d=3$.  Black stars show the energy of minimized packings and the red line is a fit to a harmonic function. Note that the simulated packings are logarithmically spaced over 10 orders of magnitude in $\varphi-\varphi_{\mathrm{J}}$ with roughly 10 systems per decade.}
\label{fig:EnergyScaling}
\end{figure}

The protocol we implemented is designed to reliably converge to an isostatic configuration. It does so by producing a series of packings with logarithmically spaced energies and excess packing fractions.  We empirically found that choosing  $n_\textrm{steps}=10$ packings per decade of packing fraction provides a reasonably high degree of success.  As shown in Figure \ref{fig:EnergyScaling}, for a quadratic contact potential as in Eq.~\eqref{eq:harmonic}, the system energy 
scales with the distance to jamming as
\begin{align}
 U \propto (\varphi - \varphi_{\mathrm{J}})^2\theta(\varphi-\varphi_{\mathrm{J}}).
\end{align}
We begin by creating a configuration at an initial packing fraction $\varphi_0 \approx 2 \varphi_{\mathrm{J}}$, and an initial estimate for the jamming density, $\tilde \varphi_0$.  We minimize the energy of this packing using the FIRE algorithm \cite{bitzek2006structural} to $U_0$, and calculate the packing fraction for the next iteration, $\varphi_1$, using the general rule
\begin{align}
  \varphi_{i+1} & = \tilde\varphi_i + \left( \varphi_i - \tilde\varphi_i \right) 10^{-1/n_\textrm{steps}}.
\end{align}
Every particle is then isotropically dilated to this new packing fraction and the system energy is minimized to $U_i$, which we then use to compute a new, better estimate for the jamming density 
\begin{align}
  \tilde\varphi_{i+1} & = \frac{\varphi_{i+1} - \varphi_i \sqrt{ U_i/U_{i-1} } }{1 - \sqrt{ U_i/U_{i-1} } }.
\end{align}
As this process evolves we see that $\varphi_i$ and $\tilde \varphi_i$ converge to $\varphi_{\mathrm{J}}$, and that $\sqrt{U_i/U_{i-1}}$ converges to $10^{-1/n_\textrm{steps}}$.  Here, we continue this procedure until $U/N\leq10^{-20}$ for $d=3$ and 4, and $U/N\leq10^{-24}$ for $d=2$.

In order to perform the energy minimization efficiently, our numerical routines make extensive use of general purpose graphical processing units (GPGPU) that are part of the University of Oregon ACISS supercomputer  (156 NVIDIA M2070).  Meeting the needed resolution between contacts and near contacts requires more precision than is offered by IEEE 754 double-precision number formats.  Our GPGPU hardware does not, however, implement IEEE 754 quadruple-precision computations.  We have thus resorted instead to implementing double-double precision algorithms, whereby each number is represented by a pair of double precision numbers, and provides 106 bits of precision in the significand (as opposed to 113 for quad precision) and 11 bits in the exponent (as opposed to 15 for quad precision).  The basic mathematical operations are based on the NVIDIA implementation of double-double precision arithmetic~\cite{NVIDIA}.

\subsection{Detailed Analysis Protocol}

The lowest energy configuration is used for subsequent analysis. Contacts and near contacts are distinguished, using a gap threshold of $10^{-11}\sigma_{ij}$, but the distinction is fairly robust to a choice of threshold within an order of magnitude of this value. Particles with $Z<d+1$ contacts are considered to be rattlers and are discarded from the rest of the analysis. Note that the rattler determination is done self-consistently, in case two or more rattlers are initially in contact with one another. After rattlers are removed, only configurations with $N_\mathrm{c}=N_{\mathrm{c}}^{\mathrm{iso}}$ are kept for the subsequent force analysis. In $d=3$ and 4, more than two-thirds of the systems met that criterion, and in $d=2$ about a quarter did so (the origin of this difference is unclear). Even though $N_\mathrm{c}$ may be off by only one or two contacts, the algorithm for extracting forces is acutely sensitive to this requirement and therefore non-isostatic configurations ought to be left out of the subsequent analysis. Further modifying these packings using a different algorithm, such as sequential linear programming~\cite{TJ10} or following the unstable modes of slightly hypostatic packings~\cite{LDW13}, may increase the yield of packings with $N_{\mathrm{c}}^{\mathrm{iso}}$ contacts, but this approach has not been attempted here.

Following the results of Sec.~\ref{sec:H}, we extract contact forces from the zero-eigenvalue eigenvector of matrix $\NN$. Because $\NN$ is sparse, the relevant part of its eigensystem can be efficiently determined with the Lanczos algorithm~\cite{La50}, as implemented in Mathematica 10~\cite{Mathematica10}. It is expected of perfectly isostatic systems that all elements of that eigenvector share the same sign, which corresponds to $f_{ij}>0$. For the vast majority of configurations, it is indeed the case. As a last check for isostaticity, the rare systems that do not meet this criterion are eliminated from the force analysis.

\subsection{Coplanarity and Weak Forces}
\begin{figure}
\includegraphics[width=0.45\columnwidth]{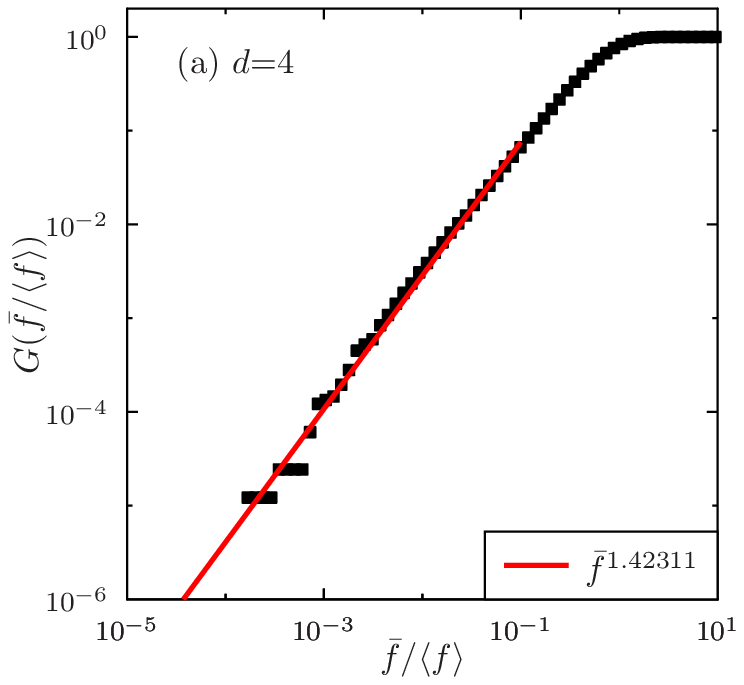}
\includegraphics[width=0.45\columnwidth]{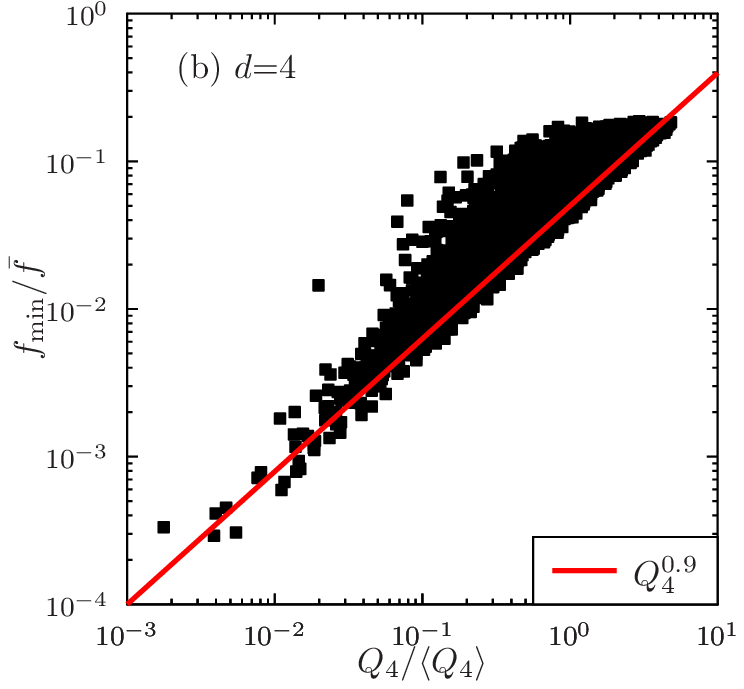}
\caption{(a) Cumulative distribution of $\bar{f}$ for particles with $Z=d+1$ contacts in systems with $N_{\mathrm{i}}=16384$ in $d=4$. The power-law scaling at weak forces is consistent with $\theta_{\mathrm{e}}$. (b) Correlation between the relative strength of the weakest force and the relative degree of coplanarity for particles with $Z=d+1$ contacts in six of these same systems.}
\label{fig:ForceScaling}
\end{figure}

In the main text, we have argued that particles with $Z=d+1$ contacts are responsible for the anomalous tail of the force distribution. The collection of particles giving rise to the anomalous scaling is further argued to have $d$  neighbors that are nearly coplanar with itself, which results in a weak resulting force (by force balance) on the $d+1$th particle and in localized soft modes. In order to disentangle the contribution of coplanarity from that of connectivity in determining what particles have an anomalous behavior, we explicitly separate the two effects below. 

First, we note that the average force on particles with $Z=d+1$
\begin{equation}
\bar{f}_i=\frac{1}{Z}\sum_{j\in\partial i}f_{ij}
\end{equation}
follows the mean-field scaling (Fig.~\ref{fig:ForceScaling}), and therefore cannot on its own capture the anomalous scaling.

Second, we validate the correlation between coplanarity and weak forces. As a measure of coplanarity, we consider the volume $Q_d$ of the parallelepiped spanned by the $d$ contact vectors that are closest to being coplanar with the center of a given particle $i$ with $Z=d+1$ contacts. In order to do so, we define a $d\times d$ matrix $\mathcal{V}$ for these $j=1\ldots d$ contact vectors as
\begin{equation}
\mathcal{V}^{j}_\alpha=r_j^{\alpha}-r_i^\alpha,
\end{equation}
where the component index $\alpha=1\ldots d$, and note that $Q_{d}=|\mathrm{det}(\mathcal{V})|$. After correcting for the inherent heterogeneity in $\bar{f}_i$, we observe a strong correlation between coplanarity and weak forces (Fig.~\ref{fig:ForceScaling}).

Joining these results with those presented in the main text confirms that the physical origin of the anomalous $\theta_\ell$ scaling is the weakest force of the particles that are most likely to buckle. Further analysis of these bucklers is left for future work.

\end{widetext}


\begin{thebibliography}{35}%
\makeatletter
\providecommand \@ifxundefined [1]{%
 \@ifx{#1\undefined}
}%
\providecommand \@ifnum [1]{%
 \ifnum #1\expandafter \@firstoftwo
 \else \expandafter \@secondoftwo
 \fi
}%
\providecommand \@ifx [1]{%
 \ifx #1\expandafter \@firstoftwo
 \else \expandafter \@secondoftwo
 \fi
}%
\providecommand \natexlab [1]{#1}%
\providecommand \enquote  [1]{``#1''}%
\providecommand \bibnamefont  [1]{#1}%
\providecommand \bibfnamefont [1]{#1}%
\providecommand \citenamefont [1]{#1}%
\providecommand \href@noop [0]{\@secondoftwo}%
\providecommand \href [0]{\begingroup \@sanitize@url \@href}%
\providecommand \@href[1]{\@@startlink{#1}\@@href}%
\providecommand \@@href[1]{\endgroup#1\@@endlink}%
\providecommand \@sanitize@url [0]{\catcode `\\12\catcode `\$12\catcode
  `\&12\catcode `\#12\catcode `\^12\catcode `\_12\catcode `\%12\relax}%
\providecommand \@@startlink[1]{}%
\providecommand \@@endlink[0]{}%
\providecommand \url  [0]{\begingroup\@sanitize@url \@url }%
\providecommand \@url [1]{\endgroup\@href {#1}{\urlprefix }}%
\providecommand \urlprefix  [0]{URL }%
\providecommand \Eprint [0]{\href }%
\providecommand \doibase [0]{http://dx.doi.org/}%
\providecommand \selectlanguage [0]{\@gobble}%
\providecommand \bibinfo  [0]{\@secondoftwo}%
\providecommand \bibfield  [0]{\@secondoftwo}%
\providecommand \translation [1]{[#1]}%
\providecommand \BibitemOpen [0]{}%
\providecommand \bibitemStop [0]{}%
\providecommand \bibitemNoStop [0]{.\EOS\space}%
\providecommand \EOS [0]{\spacefactor3000\relax}%
\providecommand \BibitemShut  [1]{\csname bibitem#1\endcsname}%
\let\auto@bib@innerbib\@empty
\bibitem [{\citenamefont {Lerner}\ \emph {et~al.}(2013)\citenamefont {Lerner},
  \citenamefont {During},\ and\ \citenamefont {Wyart}}]{LDW13}%
  \BibitemOpen
  \bibfield  {author} {\bibinfo {author} {\bibfnamefont {E.}~\bibnamefont
  {Lerner}}, \bibinfo {author} {\bibfnamefont {G.}~\bibnamefont {During}}, \
  and\ \bibinfo {author} {\bibfnamefont {M.}~\bibnamefont {Wyart}},\
  }\href@noop {} {\bibfield  {journal} {\bibinfo  {journal} {Soft Matter}\
  }\textbf {\bibinfo {volume} {9}},\ \bibinfo {pages} {8252} (\bibinfo {year}
  {2013})}\BibitemShut {NoStop}%
\bibitem [{\citenamefont {Bernal}(1959)}]{Be59}%
  \BibitemOpen
  \bibfield  {author} {\bibinfo {author} {\bibfnamefont {J.~D.}\ \bibnamefont
  {Bernal}},\ }\href@noop {} {\bibfield  {journal} {\bibinfo  {journal}
  {Nature}\ }\textbf {\bibinfo {volume} {183}},\ \bibinfo {pages} {141}
  (\bibinfo {year} {1959})},\ \bibinfo {note} {0028-0836}\BibitemShut {NoStop}%
\bibitem [{\citenamefont {Liu}\ and\ \citenamefont {Nagel}(1998)}]{LN98}%
  \BibitemOpen
  \bibfield  {author} {\bibinfo {author} {\bibfnamefont {A.~J.}\ \bibnamefont
  {Liu}}\ and\ \bibinfo {author} {\bibfnamefont {S.~R.}\ \bibnamefont
  {Nagel}},\ }\href@noop {} {\bibfield  {journal} {\bibinfo  {journal}
  {Nature}\ }\textbf {\bibinfo {volume} {396}},\ \bibinfo {pages} {21}
  (\bibinfo {year} {1998})}\BibitemShut {NoStop}%
\bibitem [{\citenamefont {van Hecke}(2010)}]{He10}%
  \BibitemOpen
  \bibfield  {author} {\bibinfo {author} {\bibfnamefont {M.}~\bibnamefont {van
  Hecke}},\ }\href@noop {} {\bibfield  {journal} {\bibinfo  {journal} {J.
  Phys.: Condens. Matt.}\ }\textbf {\bibinfo {volume} {22}},\ \bibinfo {pages}
  {033101} (\bibinfo {year} {2010})}\BibitemShut {NoStop}%
\bibitem [{\citenamefont {Torquato}\ and\ \citenamefont
  {Stillinger}(2010)}]{TS10}%
  \BibitemOpen
  \bibfield  {author} {\bibinfo {author} {\bibfnamefont {S.}~\bibnamefont
  {Torquato}}\ and\ \bibinfo {author} {\bibfnamefont {F.~H.}\ \bibnamefont
  {Stillinger}},\ }\href {\doibase 10.1103/RevModPhys.82.2633} {\bibfield
  {journal} {\bibinfo  {journal} {Rev. Mod. Phys.}\ }\textbf {\bibinfo {volume}
  {82}},\ \bibinfo {pages} {2633} (\bibinfo {year} {2010})}\BibitemShut
  {NoStop}%
\bibitem [{\citenamefont {Parisi}\ and\ \citenamefont {Zamponi}(2010)}]{PZ10}%
  \BibitemOpen
  \bibfield  {author} {\bibinfo {author} {\bibfnamefont {G.}~\bibnamefont
  {Parisi}}\ and\ \bibinfo {author} {\bibfnamefont {F.}~\bibnamefont
  {Zamponi}},\ }\href {\doibase 10.1103/RevModPhys.82.789} {\bibfield
  {journal} {\bibinfo  {journal} {Rev. Mod. Phys.}\ }\textbf {\bibinfo {volume}
  {82}},\ \bibinfo {pages} {789} (\bibinfo {year} {2010})}\BibitemShut
  {NoStop}%
\bibitem [{\citenamefont {Kurchan}\ \emph {et~al.}(2012)\citenamefont
  {Kurchan}, \citenamefont {Parisi},\ and\ \citenamefont {Zamponi}}]{KPZ12}%
  \BibitemOpen
  \bibfield  {author} {\bibinfo {author} {\bibfnamefont {J.}~\bibnamefont
  {Kurchan}}, \bibinfo {author} {\bibfnamefont {G.}~\bibnamefont {Parisi}}, \
  and\ \bibinfo {author} {\bibfnamefont {F.}~\bibnamefont {Zamponi}},\
  }\href@noop {} {\bibfield  {journal} {\bibinfo  {journal} {J. Stat. Mech.}\
  }\textbf {\bibinfo {volume} {2012}},\ \bibinfo {pages} {P10012} (\bibinfo
  {year} {2012})}\BibitemShut {NoStop}%
\bibitem [{\citenamefont {Kurchan}\ \emph {et~al.}(2013)\citenamefont
  {Kurchan}, \citenamefont {Parisi}, \citenamefont {Urbani},\ and\
  \citenamefont {Zamponi}}]{KPUZ13}%
  \BibitemOpen
  \bibfield  {author} {\bibinfo {author} {\bibfnamefont {J.}~\bibnamefont
  {Kurchan}}, \bibinfo {author} {\bibfnamefont {G.}~\bibnamefont {Parisi}},
  \bibinfo {author} {\bibfnamefont {P.}~\bibnamefont {Urbani}}, \ and\ \bibinfo
  {author} {\bibfnamefont {F.}~\bibnamefont {Zamponi}},\ }\href@noop {}
  {\bibfield  {journal} {\bibinfo  {journal} {J. Phys. Chem. B}\ }\textbf
  {\bibinfo {volume} {117}},\ \bibinfo {pages} {12979} (\bibinfo {year}
  {2013})}\BibitemShut {NoStop}%
\bibitem [{\citenamefont {Charbonneau}\ \emph
  {et~al.}(2014{\natexlab{a}})\citenamefont {Charbonneau}, \citenamefont
  {Kurchan}, \citenamefont {Parisi}, \citenamefont {Urbani},\ and\
  \citenamefont {Zamponi}}]{CKPUZ14}%
  \BibitemOpen
  \bibfield  {author} {\bibinfo {author} {\bibfnamefont {P.}~\bibnamefont
  {Charbonneau}}, \bibinfo {author} {\bibfnamefont {J.}~\bibnamefont
  {Kurchan}}, \bibinfo {author} {\bibfnamefont {G.}~\bibnamefont {Parisi}},
  \bibinfo {author} {\bibfnamefont {P.}~\bibnamefont {Urbani}}, \ and\ \bibinfo
  {author} {\bibfnamefont {F.}~\bibnamefont {Zamponi}},\ }\href@noop {}
  {\bibfield  {journal} {\bibinfo  {journal} {Nat. Comm.}\ }\textbf {\bibinfo
  {volume} {5}},\ \bibinfo {pages} {3725} (\bibinfo {year}
  {2014}{\natexlab{a}})}\BibitemShut {NoStop}%
\bibitem [{\citenamefont {Charbonneau}\ \emph
  {et~al.}(2014{\natexlab{b}})\citenamefont {Charbonneau}, \citenamefont
  {Kurchan}, \citenamefont {Parisi}, \citenamefont {Urbani},\ and\
  \citenamefont {Zamponi}}]{CKPUZ14b}%
  \BibitemOpen
  \bibfield  {author} {\bibinfo {author} {\bibfnamefont {P.}~\bibnamefont
  {Charbonneau}}, \bibinfo {author} {\bibfnamefont {J.}~\bibnamefont
  {Kurchan}}, \bibinfo {author} {\bibfnamefont {G.}~\bibnamefont {Parisi}},
  \bibinfo {author} {\bibfnamefont {P.}~\bibnamefont {Urbani}}, \ and\ \bibinfo
  {author} {\bibfnamefont {F.}~\bibnamefont {Zamponi}},\ }\href@noop {}
  {\bibfield  {journal} {\bibinfo  {journal} {J. Stat. Mech.}\ }\textbf
  {\bibinfo {volume} {2014}},\ \bibinfo {pages} {P10009} (\bibinfo {year}
  {2014}{\natexlab{b}})}\BibitemShut {NoStop}%
\bibitem [{foo({\natexlab{a}})}]{foot:2}%
  \BibitemOpen
  \href@noop {} {} \bibinfo {note} {See Appendix for an extended consideration of the excitations in a
  marginal system, for detailed discussion of isostaticity, and for numerical
  protocol details.}\BibitemShut {Stop}%
\bibitem [{\citenamefont {Wyart}\ \emph {et~al.}(2005)\citenamefont {Wyart},
  \citenamefont {Silbert}, \citenamefont {Nagel},\ and\ \citenamefont
  {Witten}}]{WSNW05}%
  \BibitemOpen
  \bibfield  {author} {\bibinfo {author} {\bibfnamefont {M.}~\bibnamefont
  {Wyart}}, \bibinfo {author} {\bibfnamefont {L.}~\bibnamefont {Silbert}},
  \bibinfo {author} {\bibfnamefont {S.}~\bibnamefont {Nagel}}, \ and\ \bibinfo
  {author} {\bibfnamefont {T.}~\bibnamefont {Witten}},\ }\href@noop {}
  {\bibfield  {journal} {\bibinfo  {journal} {Phys. Rev. E}\ }\textbf {\bibinfo
  {volume} {72}},\ \bibinfo {pages} {051306} (\bibinfo {year}
  {2005})}\BibitemShut {NoStop}%
\bibitem [{\citenamefont {Brito}\ and\ \citenamefont {Wyart}(2009)}]{Brito09}%
  \BibitemOpen
  \bibfield  {author} {\bibinfo {author} {\bibfnamefont {C.}~\bibnamefont
  {Brito}}\ and\ \bibinfo {author} {\bibfnamefont {M.}~\bibnamefont {Wyart}},\
  }\href {\doibase 10.1063/1.3157261} {\bibfield  {journal} {\bibinfo
  {journal} {J. Chem. Phys.}\ }\textbf {\bibinfo {volume} {131}},\ \bibinfo
  {eid} {024504} (\bibinfo {year} {2009})}\BibitemShut {NoStop}%
\bibitem [{\citenamefont {Wyart}(2012)}]{Wy12}%
  \BibitemOpen
  \bibfield  {author} {\bibinfo {author} {\bibfnamefont {M.}~\bibnamefont
  {Wyart}},\ }\href@noop {} {\bibfield  {journal} {\bibinfo  {journal} {Phys.
  Rev. Lett.}\ }\textbf {\bibinfo {volume} {109}},\ \bibinfo {pages} {125502}
  (\bibinfo {year} {2012})}\BibitemShut {NoStop}%
\bibitem [{\citenamefont {DeGiuli}\ \emph
  {et~al.}(2014{\natexlab{a}})\citenamefont {DeGiuli}, \citenamefont
  {Laversanne-Finot}, \citenamefont {D\"uring}, \citenamefont {Lerner},\ and\
  \citenamefont {Wyart}}]{DeGiuli14}%
  \BibitemOpen
  \bibfield  {author} {\bibinfo {author} {\bibfnamefont {E.}~\bibnamefont
  {DeGiuli}}, \bibinfo {author} {\bibfnamefont {A.}~\bibnamefont
  {Laversanne-Finot}}, \bibinfo {author} {\bibfnamefont {G.~A.}\ \bibnamefont
  {D\"uring}}, \bibinfo {author} {\bibfnamefont {E.}~\bibnamefont {Lerner}}, \
  and\ \bibinfo {author} {\bibfnamefont {M.}~\bibnamefont {Wyart}},\
  }\href@noop {} {\bibfield  {journal} {\bibinfo  {journal} {Soft Matter}\
  }\textbf {\bibinfo {volume} {10}},\ \bibinfo {pages} {5628} (\bibinfo {year}
  {2014}{\natexlab{a}})}\BibitemShut {NoStop}%
\bibitem [{\citenamefont {DeGiuli}\ \emph
  {et~al.}(2014{\natexlab{b}})\citenamefont {DeGiuli}, \citenamefont {Lerner},
  \citenamefont {Brito},\ and\ \citenamefont {Wyart}}]{DLBW14}%
  \BibitemOpen
  \bibfield  {author} {\bibinfo {author} {\bibfnamefont {E.}~\bibnamefont
  {DeGiuli}}, \bibinfo {author} {\bibfnamefont {E.}~\bibnamefont {Lerner}},
  \bibinfo {author} {\bibfnamefont {C.}~\bibnamefont {Brito}}, \ and\ \bibinfo
  {author} {\bibfnamefont {M.}~\bibnamefont {Wyart}},\ }\href@noop {}
  {\bibfield  {journal} {\bibinfo  {journal} {Proc. Nat. Acad. Sci., U.S.A.}\
  }\textbf {\bibinfo {volume} {111}},\ \bibinfo {pages} {17054} (\bibinfo
  {year} {2014}{\natexlab{b}})}\BibitemShut {NoStop}%
\bibitem [{\citenamefont {M\"uller}\ and\ \citenamefont {Wyart}(2015)}]{MW14}%
  \BibitemOpen
  \bibfield  {author} {\bibinfo {author} {\bibfnamefont {M.}~\bibnamefont
  {M\"uller}}\ and\ \bibinfo {author} {\bibfnamefont {M.}~\bibnamefont
  {Wyart}},\ }\href@noop {} {\bibfield  {journal} {\bibinfo  {journal} {Ann.
  Rev. Cond. Matt. Phys.}\ }\textbf {\bibinfo {volume} {6}},\ \bibinfo {pages}
  {9.1} (\bibinfo {year} {2015})}\BibitemShut {NoStop}%
\bibitem [{foo({\natexlab{b}})}]{foot:1}%
  \BibitemOpen
  \href@noop {} {} \bibinfo {note} {$Z(h)$ is also the
  integral of the radial pair distribution function, i.e., $Z(h)=\r
  \int_0^{(1+h)\sigma} g(r) d\mathbf{r}$, where $\r$ is the system density. Its
  critical decay is therefore $g(h)\sim h^{-\gamma}$.}\BibitemShut {Stop}%
\bibitem [{\citenamefont {Lerner}\ \emph {et~al.}(2014)\citenamefont {Lerner},
  \citenamefont {DeGiuli}, \citenamefont {D{\"u}ring},\ and\ \citenamefont
  {Wyart}}]{LDDW14}%
  \BibitemOpen
  \bibfield  {author} {\bibinfo {author} {\bibfnamefont {E.}~\bibnamefont
  {Lerner}}, \bibinfo {author} {\bibfnamefont {E.}~\bibnamefont {DeGiuli}},
  \bibinfo {author} {\bibfnamefont {G.}~\bibnamefont {D{\"u}ring}}, \ and\
  \bibinfo {author} {\bibfnamefont {M.}~\bibnamefont {Wyart}},\ }\href@noop {}
  {\bibfield  {journal} {\bibinfo  {journal} {Soft Matter}\ } (\bibinfo {year}
  {2014})}\BibitemShut {NoStop}%
\bibitem [{\citenamefont {Skoge}\ \emph {et~al.}(2006)\citenamefont {Skoge},
  \citenamefont {Donev}, \citenamefont {Stillinger},\ and\ \citenamefont
  {Torquato}}]{SDST06}%
  \BibitemOpen
  \bibfield  {author} {\bibinfo {author} {\bibfnamefont {M.}~\bibnamefont
  {Skoge}}, \bibinfo {author} {\bibfnamefont {A.}~\bibnamefont {Donev}},
  \bibinfo {author} {\bibfnamefont {F.~H.}\ \bibnamefont {Stillinger}}, \ and\
  \bibinfo {author} {\bibfnamefont {S.}~\bibnamefont {Torquato}},\ }\href@noop
  {} {\bibfield  {journal} {\bibinfo  {journal} {Phys. Rev. E}\ }\textbf
  {\bibinfo {volume} {74}},\ \bibinfo {eid} {041127} (\bibinfo {year}
  {2006})}\BibitemShut {NoStop}%
\bibitem [{\citenamefont {Goodrich}\ \emph {et~al.}(2012)\citenamefont
  {Goodrich}, \citenamefont {Liu},\ and\ \citenamefont {Nagel}}]{GLN12}%
  \BibitemOpen
  \bibfield  {author} {\bibinfo {author} {\bibfnamefont {C.~P.}\ \bibnamefont
  {Goodrich}}, \bibinfo {author} {\bibfnamefont {A.~J.}\ \bibnamefont {Liu}}, \
  and\ \bibinfo {author} {\bibfnamefont {S.~R.}\ \bibnamefont {Nagel}},\
  }\href@noop {} {\bibfield  {journal} {\bibinfo  {journal} {Phys. Rev. Lett.}\
  }\textbf {\bibinfo {volume} {109}},\ \bibinfo {pages} {095704} (\bibinfo
  {year} {2012})}\BibitemShut {NoStop}%
\bibitem [{\citenamefont {Charbonneau}\ \emph {et~al.}(2012)\citenamefont
  {Charbonneau}, \citenamefont {Corwin}, \citenamefont {Parisi},\ and\
  \citenamefont {Zamponi}}]{CCPZ12}%
  \BibitemOpen
  \bibfield  {author} {\bibinfo {author} {\bibfnamefont {P.}~\bibnamefont
  {Charbonneau}}, \bibinfo {author} {\bibfnamefont {E.~I.}\ \bibnamefont
  {Corwin}}, \bibinfo {author} {\bibfnamefont {G.}~\bibnamefont {Parisi}}, \
  and\ \bibinfo {author} {\bibfnamefont {F.}~\bibnamefont {Zamponi}},\
  }\href@noop {} {\bibfield  {journal} {\bibinfo  {journal} {Phys. Rev. Lett.}\
  }\textbf {\bibinfo {volume} {109}},\ \bibinfo {pages} {205501} (\bibinfo
  {year} {2012})}\BibitemShut {NoStop}%
\bibitem [{\citenamefont {O'Hern}\ \emph {et~al.}(2002)\citenamefont {O'Hern},
  \citenamefont {Langer}, \citenamefont {Liu},\ and\ \citenamefont
  {Nagel}}]{OLLN02}%
  \BibitemOpen
  \bibfield  {author} {\bibinfo {author} {\bibfnamefont {C.~S.}\ \bibnamefont
  {O'Hern}}, \bibinfo {author} {\bibfnamefont {S.~A.}\ \bibnamefont {Langer}},
  \bibinfo {author} {\bibfnamefont {A.~J.}\ \bibnamefont {Liu}}, \ and\
  \bibinfo {author} {\bibfnamefont {S.~R.}\ \bibnamefont {Nagel}},\ }\href
  {\doibase 10.1103/PhysRevLett.88.075507} {\bibfield  {journal} {\bibinfo
  {journal} {Phys. Rev. Lett.}\ }\textbf {\bibinfo {volume} {88}},\ \bibinfo
  {pages} {075507} (\bibinfo {year} {2002})}\BibitemShut {NoStop}%
\bibitem [{\citenamefont {Donev}\ \emph {et~al.}(2005)\citenamefont {Donev},
  \citenamefont {Torquato},\ and\ \citenamefont {Stillinger}}]{DTS05}%
  \BibitemOpen
  \bibfield  {author} {\bibinfo {author} {\bibfnamefont {A.}~\bibnamefont
  {Donev}}, \bibinfo {author} {\bibfnamefont {S.}~\bibnamefont {Torquato}}, \
  and\ \bibinfo {author} {\bibfnamefont {F.~H.}\ \bibnamefont {Stillinger}},\
  }\href {\doibase 10.1103/PhysRevE.71.011105} {\bibfield  {journal} {\bibinfo
  {journal} {Phys. Rev. E}\ }\textbf {\bibinfo {volume} {71}},\ \bibinfo {eid}
  {011105} (\bibinfo {year} {2005})}\BibitemShut {NoStop}%
\bibitem [{\citenamefont {Hopkins}\ \emph {et~al.}(2013)\citenamefont
  {Hopkins}, \citenamefont {Stillinger},\ and\ \citenamefont
  {Torquato}}]{HST13}%
  \BibitemOpen
  \bibfield  {author} {\bibinfo {author} {\bibfnamefont {A.~B.}\ \bibnamefont
  {Hopkins}}, \bibinfo {author} {\bibfnamefont {F.~H.}\ \bibnamefont
  {Stillinger}}, \ and\ \bibinfo {author} {\bibfnamefont {S.}~\bibnamefont
  {Torquato}},\ }\href@noop {} {\bibfield  {journal} {\bibinfo  {journal}
  {Phys. Rev. E}\ }\textbf {\bibinfo {volume} {88}},\ \bibinfo {pages} {022205}
  (\bibinfo {year} {2013})}\BibitemShut {NoStop}%
\bibitem [{\citenamefont {Torquato}\ and\ \citenamefont {Jiao}(2010)}]{TJ10}%
  \BibitemOpen
  \bibfield  {author} {\bibinfo {author} {\bibfnamefont {S.}~\bibnamefont
  {Torquato}}\ and\ \bibinfo {author} {\bibfnamefont {Y.}~\bibnamefont
  {Jiao}},\ }\href@noop {} {\bibfield  {journal} {\bibinfo  {journal} {Phys.
  Rev. E}\ }\textbf {\bibinfo {volume} {82}},\ \bibinfo {pages} {061302}
  (\bibinfo {year} {2010})}\BibitemShut {NoStop}%
\bibitem [{\citenamefont {Morse}\ and\ \citenamefont
  {Corwin}(2014)}]{morse2014geometric}%
  \BibitemOpen
  \bibfield  {author} {\bibinfo {author} {\bibfnamefont {P.~K.}\ \bibnamefont
  {Morse}}\ and\ \bibinfo {author} {\bibfnamefont {E.~I.}\ \bibnamefont
  {Corwin}},\ }\href@noop {} {\bibfield  {journal} {\bibinfo  {journal} {Phys.
  Rev. Lett.}\ }\textbf {\bibinfo {volume} {112}},\ \bibinfo {pages} {115701}
  (\bibinfo {year} {2014})}\BibitemShut {NoStop}%
\bibitem [{\citenamefont {NVIDIA}(2013)}]{NVIDIA}%
  \BibitemOpen
  \bibfield  {author} {\bibinfo {author} {\bibnamefont {NVIDIA}},\ }\href@noop
  {} {\enquote {\bibinfo {title} {Double-double precision arithmetic},}\
  }\bibinfo {howpublished}
  {\url{https://developer.nvidia.com/rdp/assets/double-double-precision-arithm%
etic}} (\bibinfo {year} {2013}),\ \bibinfo {note} {last accessed November 14,
  2014}\BibitemShut {NoStop}%
\bibitem [{\citenamefont {Bitzek}\ \emph {et~al.}(2006)\citenamefont {Bitzek},
  \citenamefont {Koskinen}, \citenamefont {G\"ahler}, \citenamefont {Moseler},\
  and\ \citenamefont {Gumbsch}}]{bitzek2006structural}%
  \BibitemOpen
  \bibfield  {author} {\bibinfo {author} {\bibfnamefont {E.}~\bibnamefont
  {Bitzek}}, \bibinfo {author} {\bibfnamefont {P.}~\bibnamefont {Koskinen}},
  \bibinfo {author} {\bibfnamefont {F.}~\bibnamefont {G\"ahler}}, \bibinfo
  {author} {\bibfnamefont {M.}~\bibnamefont {Moseler}}, \ and\ \bibinfo
  {author} {\bibfnamefont {P.}~\bibnamefont {Gumbsch}},\ }\href {\doibase
  10.1103/PhysRevLett.97.170201} {\bibfield  {journal} {\bibinfo  {journal}
  {Phys. Rev. Lett.}\ }\textbf {\bibinfo {volume} {97}},\ \bibinfo {pages}
  {170201} (\bibinfo {year} {2006})}\BibitemShut {NoStop}%
\bibitem [{\citenamefont {Jacquin}\ \emph {et~al.}(2011)\citenamefont
  {Jacquin}, \citenamefont {Berthier},\ and\ \citenamefont {Zamponi}}]{JBZ11}%
  \BibitemOpen
  \bibfield  {author} {\bibinfo {author} {\bibfnamefont {H.}~\bibnamefont
  {Jacquin}}, \bibinfo {author} {\bibfnamefont {L.}~\bibnamefont {Berthier}}, \
  and\ \bibinfo {author} {\bibfnamefont {F.}~\bibnamefont {Zamponi}},\
  }\href@noop {} {\bibfield  {journal} {\bibinfo  {journal} {Phys. Rev. Lett.}\
  }\textbf {\bibinfo {volume} {106}},\ \bibinfo {pages} {135702} (\bibinfo
  {year} {2011})}\BibitemShut {NoStop}%
\bibitem [{\citenamefont {Perera}\ and\ \citenamefont
  {Harrowell}(1998)}]{perera:1998}%
  \BibitemOpen
  \bibfield  {author} {\bibinfo {author} {\bibfnamefont {D.~N.}\ \bibnamefont
  {Perera}}\ and\ \bibinfo {author} {\bibfnamefont {P.}~\bibnamefont
  {Harrowell}},\ }\href@noop {} {\bibfield  {journal} {\bibinfo  {journal}
  {Phys. Rev. Lett.}\ }\textbf {\bibinfo {volume} {81}},\ \bibinfo {pages}
  {120} (\bibinfo {year} {1998})}\BibitemShut {NoStop}%
\bibitem [{\citenamefont {van Meel}\ \emph {et~al.}(2009)\citenamefont {van
  Meel}, \citenamefont {Charbonneau}, \citenamefont {Fortini},\ and\
  \citenamefont {Charbonneau}}]{VCFC09}%
  \BibitemOpen
  \bibfield  {author} {\bibinfo {author} {\bibfnamefont {J.~A.}\ \bibnamefont
  {van Meel}}, \bibinfo {author} {\bibfnamefont {B.}~\bibnamefont
  {Charbonneau}}, \bibinfo {author} {\bibfnamefont {A.}~\bibnamefont
  {Fortini}}, \ and\ \bibinfo {author} {\bibfnamefont {P.}~\bibnamefont
  {Charbonneau}},\ }\href@noop {} {\bibfield  {journal} {\bibinfo  {journal}
  {Phys. Rev. E}\ }\textbf {\bibinfo {volume} {80}},\ \bibinfo {pages} {061110}
  (\bibinfo {year} {2009})}\BibitemShut {NoStop}%
\bibitem [{\citenamefont {Lanczos}(1950)}]{La50}%
  \BibitemOpen
  \bibfield  {author} {\bibinfo {author} {\bibfnamefont {C.}~\bibnamefont
  {Lanczos}},\ }\href@noop {} {\bibfield  {journal} {\bibinfo  {journal} {J.
  Res. Nat. Bur. Stand.}\ }\textbf {\bibinfo {volume} {45}},\ \bibinfo {pages}
  {255} (\bibinfo {year} {1950})}\BibitemShut {NoStop}%
\bibitem [{Mat(2014)}]{Mathematica10}%
  \BibitemOpen
  \href@noop {} {\emph {\bibinfo {title} {``Mathematica, Version 10.0"}}}\
  (\bibinfo  {publisher} {Wolfram Research, Inc.},\ \bibinfo {address}
  {Champaign},\ \bibinfo {year} {2014})\BibitemShut {NoStop}%
\bibitem [{\citenamefont {Urbani}\ and\ \citenamefont {Biroli}(2014)}]{UB14}%
  \BibitemOpen
  \bibfield  {author} {\bibinfo {author} {\bibfnamefont {P.}~\bibnamefont
  {Urbani}}\ and\ \bibinfo {author} {\bibfnamefont {G.}~\bibnamefont
  {Biroli}},\ }\href@noop {} {\bibfield  {journal} {\bibinfo  {journal} {{\tt
  arXiv:1410.4523}}\ } (\bibinfo {year} {2014})}\BibitemShut {NoStop}%
\end{thebibliography}
\end{document}